\documentclass[aps,a4,twocolumn,showpacs]{revtex4-1}
\usepackage{amsmath}
\usepackage{latexsym}
\usepackage{float}
\usepackage{amssymb}
\usepackage{graphicx}
\usepackage{epsfig}

\begin{document}

\title{Estimation of Persistence Lengths of Semiflexible Polymers: Insight from Simulations}
\author{Hsiao-Ping Hsu$^a$, Wolfgang Paul$^b$, and Kurt Binder$^a$}

\affiliation{$^a$Institut f\"ur Physik, Johannes Gutenberg-Universit\"at Mainz,\\
 Staudinger Weg 7, D-55099 Mainz, Germany \\
$^b$Theoretische Physik, Martin Luther Universit\"at \\
Halle-Wittenberg, von Seckendorffplatz 1, 06120 Halle, Germany}

\date{\today}

\begin{abstract}
The persistence length of macromolecules is one of their basic characteristics, describing their 
intrinsic local stiffness. However, it is difficult to extract this length from physical properties 
of the polymers, different recipes may give answers that disagree with each
other. Monte Carlo simulations are used to elucidate this problem, giving a comparative discussion of 
two lattice models, the self-avoiding walk model extended by a bond bending energy, and bottle-brush 
polymers described by the bond fluctuation model. The conditions
are discussed under which a description of such macromolecules by 
Kratky-Porod worm-like chains holds, 
and the question to what extent the persistence length depends on external conditions 
(such as solvent quality) is considered. The scattering function of semiflexible
polymers is discussed in detail, a comparison to various analytic treatments is given, and an outlook 
to experimental work is presented.
\end{abstract}

%\keywords{Monte Carlo method, persistence length, scaling, 
%semiflexible polymers, structure factor}

\maketitle
\newpage
\section{Introduction}
Flexibility of chain molecules (or lack of flexibility, respectively) is one of their most basic 
general properties~\cite{1,2,3,4,5}. It affects the use of macromolecules as building entities of 
soft materials, and controls some aspects of the functions of biopolymers in a biological context. 
Thus, it is important to understand its origin in terms of the macromolecular chemical architecture, 
and the extent to which it depends on external conditions (temperature, solvent quality if the polymer 
is in solution, as well as polymer concentration), and one therefore needs to be able to characterize 
macromolecular flexibility or stiffness precisely.
The quantity that is supposed to describe the local intrinsic stiffness of a polymer is termed 
``persistence length'' and often it is introduced (e.g.~\cite{4,5}) as a length describing the 
exponential decay of orientational correlations of segments with the length of the piece of the chain 
separating them. Thus, let us consider a linear macromolecule composed of segments vectors
\{$\vec{a}_i$, $i=1, \cdots, N)$\}, all having the same bond length 
$\ell_b \,(\langle \vec{a}^2_i \rangle=\ell^2_b$, if we wish to allow for thermal fluctuations of 
the length of these segments). Then it is assumed that the correlation of 
two segments $i$, $j$, that are $s=|i-j|$ steps along the chain apart, varies as
\begin{equation} \label{eq1}
\langle \cos \theta (s) \rangle  = 
\langle \vec{a}_i \cdot \vec{a}_j \rangle/
\langle a^2_i \rangle =\exp(-s \ell_b/\ell_p) \, , 
\quad s \rightarrow \infty ,
\end{equation}
where $\ell_p$ is the persistence length.

\begin{figure}
\begin{center}
% for a multi-line caption
%\onelinecaptionstrue
% for a one-line caption
%\includegraphics{}
(a)\includegraphics[scale=0.32,angle=270]{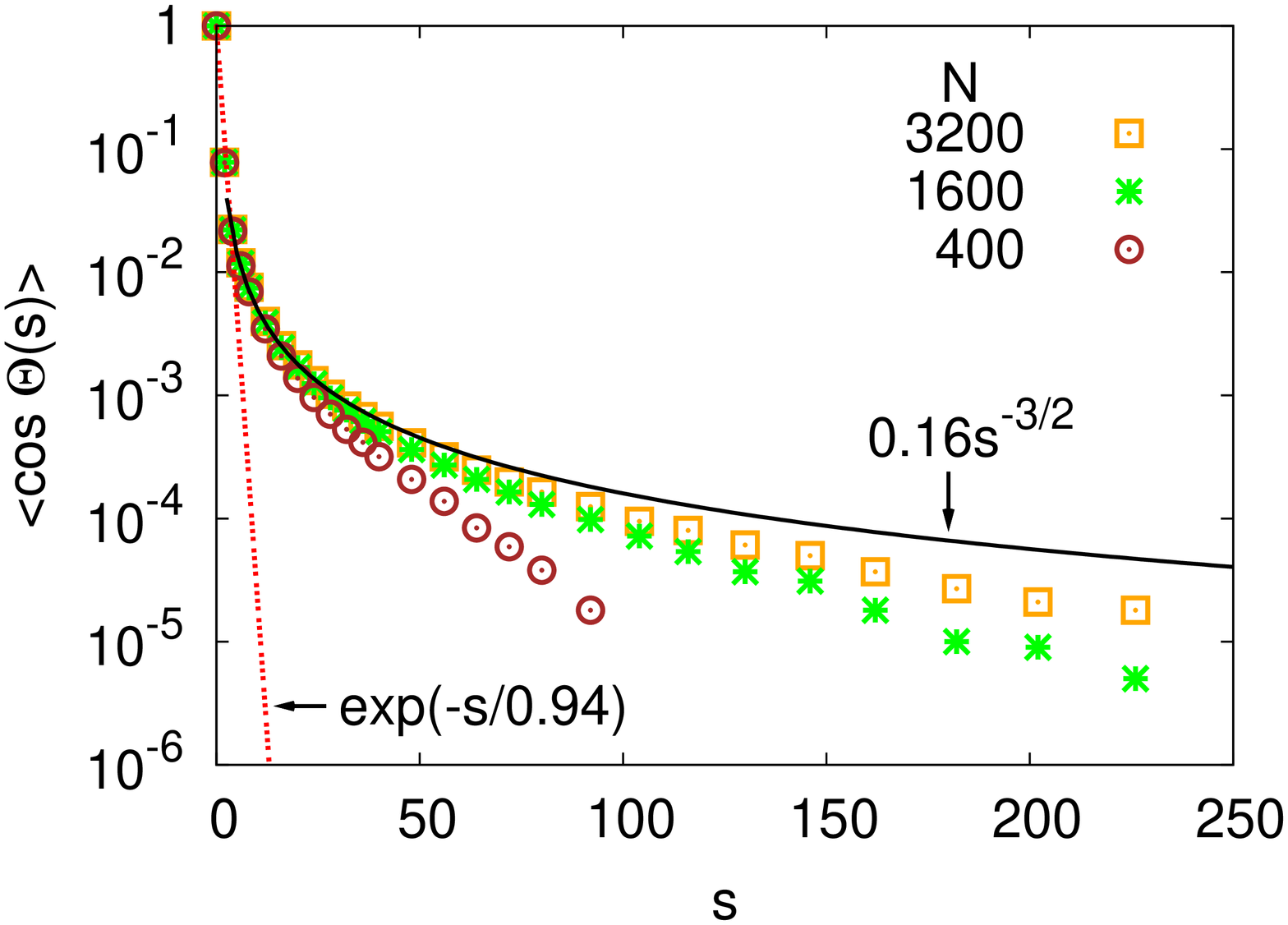}\hspace{0.4cm}
(b)\includegraphics[scale=0.32,angle=270]{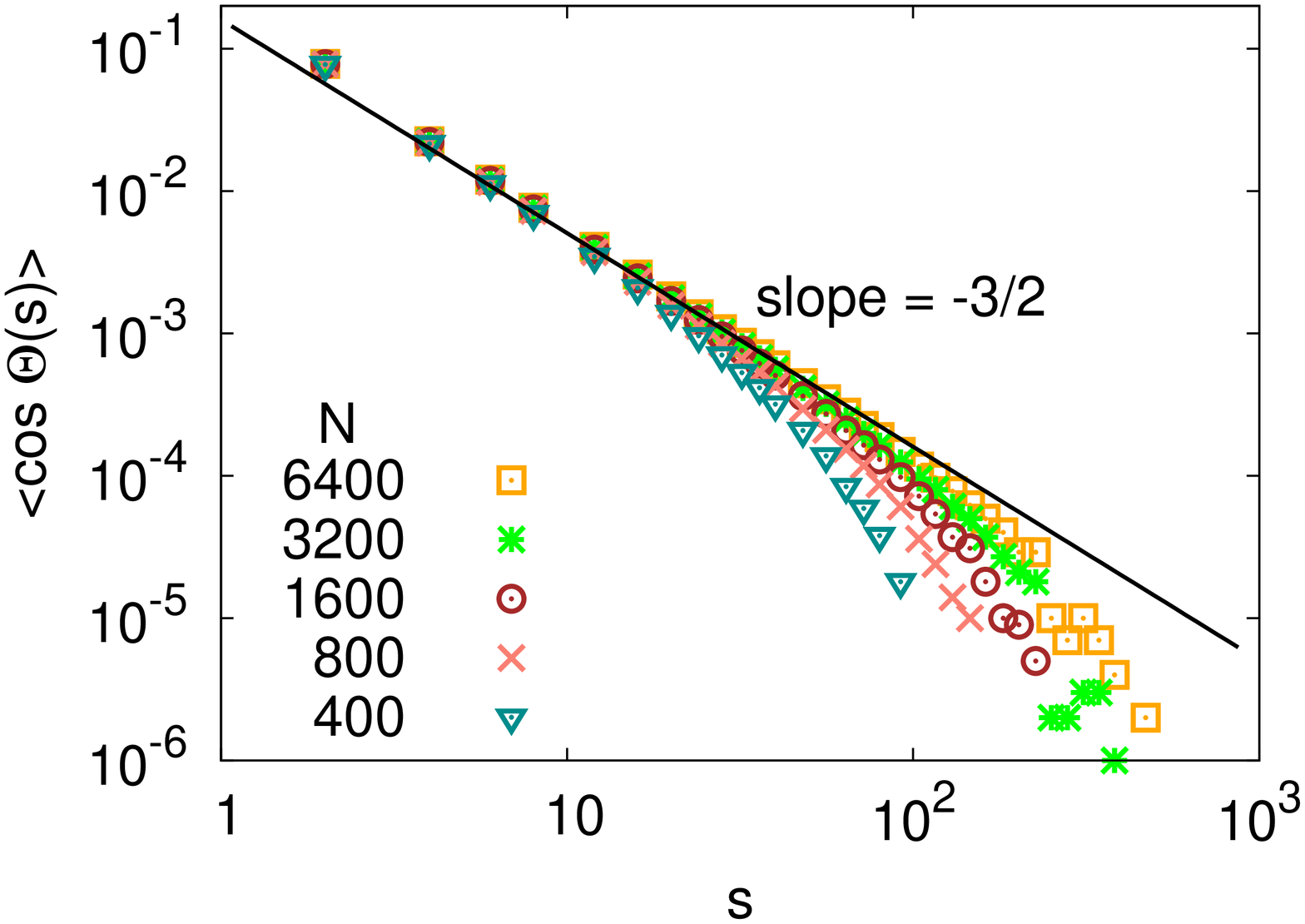}\\
\caption{\label{fig1} Semi-log plot (a) and Log-log plot (b) of
$\langle \cos \theta (s) \rangle$ versus $s$ as obtained from Monte
Carlo simulations (as described in~\cite{11}) using the pruned-enriched
Rosenbluth method
(PERM algorithm~\cite{12}) for a self-avoiding walk with nearest-neighbor
attraction $\varepsilon$, under
Theta point conditions. The full curve in (a) and straight line in (b)
represents the relation $\langle
\cos \theta (s) \rangle=0.16 s^{-3/2}$.}
\end{center}
\end{figure}

In fact, Eq.~(\ref{eq1}) holds for models of linear polymer chains that 
strictly follow Gaussian
statistics (for large distances between monomeric units), however, 
Eq.~(\ref{eq1}) is not true
for real polymers, irrespective of the considered conditions: for dilute 
solutions and good solvent
conditions one rather finds a power law behavior~\cite{6} 
\begin{equation} \label{eq2}
\langle \cos \theta (s) \rangle \propto s^{- \beta} \,, \quad
\beta=2 (1 - \nu) \, , \quad 1 \ll s \ll N \,.
\end{equation}
Here $\nu$ is the well-know Flory exponent, describing the scaling of 
the end-to-end distance
$\vec{R}= \sum\limits_{i=1}^{N} \, \vec{a}_i$ with the number  $N$ of segments, 
$\langle R^2 \rangle \propto N^{2 \nu}$, with $\nu \approx 3/5$ 
(more precisely~\cite{7}, $\nu=0.588$) 
in $d=3$ dimensions~\cite{1,2,3,4,5}. Polymer chains in 
dense melts do show a scaling of 
the end-to-end distance as predicted by Gaussian statistics, 
$\langle R^2 \rangle \propto N$ 
(i.e., $\nu$ takes the mean-field value $\nu_{MF}=1/2)$, and hence it 
was widely believed, that 
Eq.~(\ref{eq1}) is useful for polymer chains under melt conditions. 
However, recent analytical and 
numerical work~\cite{8,9} has shown that this assertion is 
completely wrong, and there also holds a power law
decay, though with a different exponent,
\begin{equation} \label{eq3}
\langle \cos \theta (s) \rangle \propto s^{-3/2}, \quad 1 \ll s \ll N.
\end{equation}
More recently, it was also found by approximate analytical 
arguments~\cite{10}, and verified in 
extensive simulations~\cite{11} that Eq.~(\ref{eq3}) also holds for 
chains in dilute solutions at 
the Theta point. In practice, since asymptotic power laws such as 
Eqs.~(\ref{eq2}), (\ref{eq3}) hold only 
in the intermediate regime $1 \ll s \ll N$ and hence one must consider 
the limit $N \rightarrow \infty$, 
one easily could be misled if data for $\langle \cos \theta (s) \rangle$ 
are considered for 
insufficiently long chains. As an example Fig.~\ref{fig1} presents 
simulation results for the simple 
self-avoiding walk (SAW) model on the simple cubic (sc) lattice, where an 
attractive energy $\varepsilon$ 
between neighboring occupied sites (representing the effective 
monomers of the chain) occurs and the 
temperature is chosen as $k_BT/\varepsilon=3.717$ which is known to 
reproduce Theta point conditions for 
this model~\cite{12}. One can see clearly that the data for 
$N \rightarrow \infty$ and
$s \geq 10$ do approach Eq.~(\ref{eq3}), but for finite $N$ 
systematic deviations from Eq.~(\ref{eq3}) 
clearly are visible already for $s=N/10$. On the semi-log plot, 
for rather short chains one might
be tempted to apply a fit of an exponential decay proportional 
to $\exp(-s \ell_b/ \ell_p)$ to the data 
for rather large $s$, but resulting estimates for $\ell_p/\ell_b$ are 
not meaningful at all: for the 
considered model, the chain is fully flexible, any reasonable estimate 
for $\ell_p/\ell_b$ that 
describes the local intrinsic stiffness of the chain should be
(i) of order unity 
{(see Fig.~\ref{fig1}a, $\ell_p/\ell_b\approx 0.94$)}, 
and (ii) independent of $N$. Both conditions 
are dramatically violated, of course, 
if estimates for $\ell_p/\ell_b$ were extracted from fits to an 
exponential decay in this way.

\begin{figure}
\begin{center}
% for a multi-line caption
%\onelinecaptionstrue
% for a one-line caption
%\includegraphics{}
(a)\includegraphics[scale=0.32,angle=270]{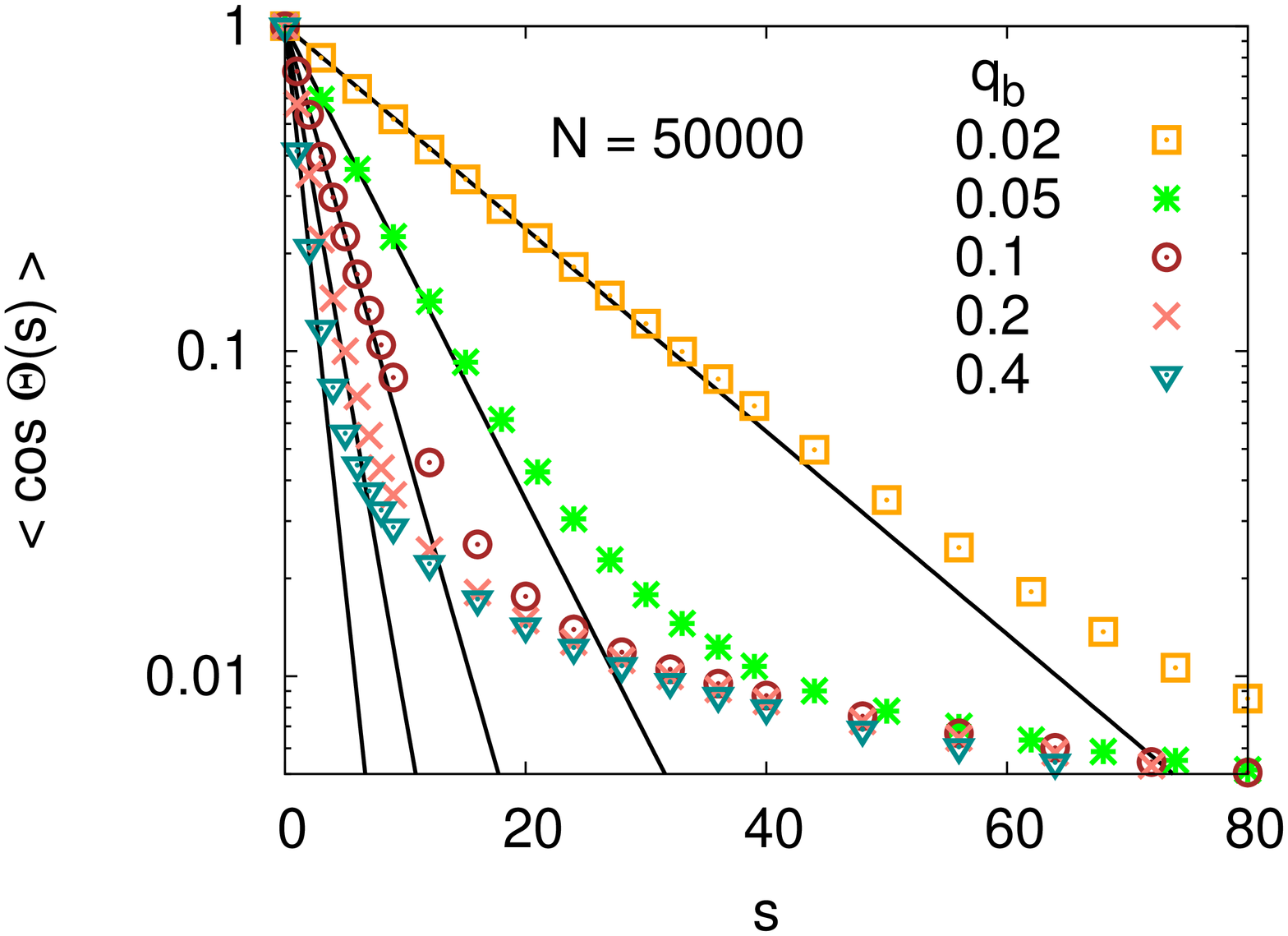}\hspace{0.4cm}
(b)\includegraphics[scale=0.32,angle=270]{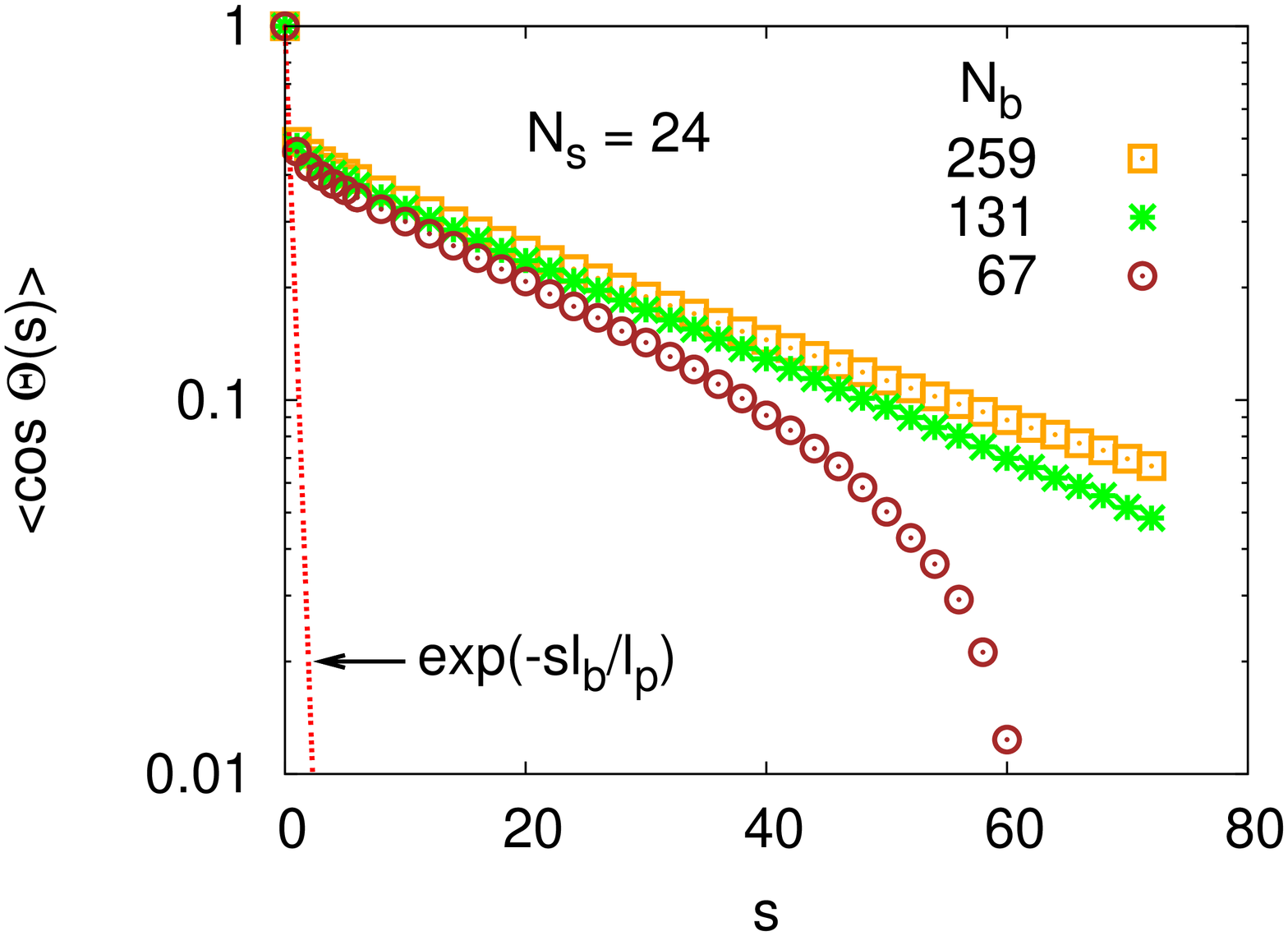}\\
\caption{\label{fig2} Semi-log of $\langle \cos \theta (s) \rangle$
vs.~$s$ for a semiflexible version of a SAW model on the sc lattice (a),
cf. text, and the bond-fluctuation model of bottle-brush polymers under
very good solvent conditions~\cite{11,17} (b). Part (a) refers to the
chains of length $N=50000$, and several choices of
the parameter $q_b=\exp (-\varepsilon_b/k_BT)$
controlling the chain stiffness, namely $q_b=0.4$, 0.2$, 0.1, 0.05$
and $0.02$. 
{Using Eq.~(\ref{eq4}), the straight lines indicate
the exponential decay $\exp(-s\ell_b/\ell_p)$ for the choices of $q_b$.}
Part (b) refers to the case of bottle-brush polymers where
every effective monomer of the backbone has one side chain of length
$N_s=24$ grafted to it, and several choices of
backbone chain length $N_b$.
Here $\ell_p/\ell_b=-1/\ln(\langle \cos \theta(1) \rangle)$ has been
extracted from the chain backbone only.}
\end{center}
\end{figure}

\begin{figure}
\begin{center}
% for a multi-line caption
%\onelinecaptionstrue
% for a one-line caption
%\includegraphics{}
\includegraphics[scale=0.25,angle=0]{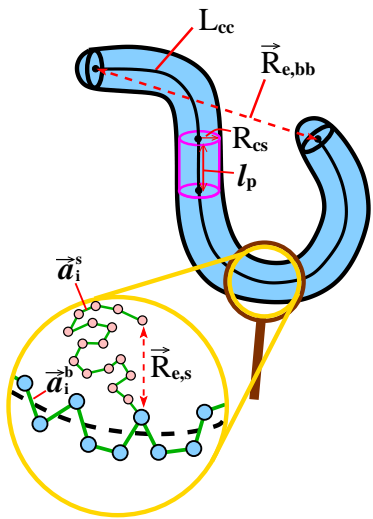}\hspace{0.4cm}
\caption{\label{fig3} Sketch of the multiple length scales that
one may define for bottle-brush polymers
(schematic): While on scales where one resolves the effective
monomers of both the backbone and the side chains, the correlations
of backbone vectors $\vec{a}_i^b$ (which have length $\ell_b$) and
side chain bond vectors
$\vec{a}^s_i$ (which have length $\ell_s$) can be studied, as well
as end-to-end distances $\vec{R}_{e,s}$ of side chains and
of the backbone $\vec{R}_{e,bb}$ (and the corresponding gyration
radii). In the bond-fluctuation model of Fig.~\ref{fig2},
for simplicity no chemical difference between backbone and monomer
was considered, so $\ell_b=\ell_s$ was chosen. The
``microscopic'' contour length of the backbone then is
$L_b=N_b\ell_b$, if the backbone has $N_b$ bonds. On a coarse-grained
level the bottle-brush resembles a worm-like chain of thickness
(cross-sectional radius) $R_{cs}$ and contour length
$L_{cc}< L_b$, which is locally straight on the scale of the
persistence length $\ell_p$.}
\end{center}
\end{figure}

Since the intrinsic stiffness of a chain is a local property of a 
macromolecule, one might alternatively 
try the recipe to either fit Eq.~(\ref{eq1}) in the regime of small $s$ 
to the data, or assume that 
Eq.~(\ref{eq1}) holds for $s=1$ already and hence
\begin{equation} \label{eq4}
\ell_p/\ell_b= -1 / \ln ( \langle \cos \theta (1) \rangle ) \quad.
\end{equation}
This recipe works in simple cases, such as the SAW model where an energy 
$\varepsilon_b$ associated with 
bond bending is added (every kink of the walk by $\pm$ 90$^o$ on the sc 
lattice costs $\varepsilon_b$), 
see Fig.~\ref{fig2}, but it fails for molecules with more complex chemical 
architecture, such as bottle-brush 
molecules~\cite{13,14,15,16}. The dramatic failure of Eq.~(\ref{eq4}) for 
bottle-brush polymers is understood 
in terms of their multiscale structure
(Fig.~\ref{fig3}): The side chains lead to a stiffness of the backbone on 
a mesoscopic scale, even if on the 
local scale of nearest-neighbor bonds the backbone is still rather 
flexible. The question of understanding 
this stiffening of bottle-brush polymers because of their grafted linear 
side chains~\cite{11,13,14,15,16,17,18,19,20,21,22,23,24,25,26,27,28,29,30,31,32,33,34,35,36,37,38} or 
grafted branched objects~\cite{39,40,41,42} is an issue of longstanding 
debate in the literature.

Complex polymer architecture is only one out of many reasons which 
make the analysis of bond orientational 
correlations based on Eqs.~(\ref{eq1}) or (\ref{eq4}) problematic. 
In dilute solutions we expect that a 
nontrivial crossover occurs when the solvent quality is marginal, 
i.e.~close to the Theta point a large 
size $\xi_T$ of ``thermal blobs''~\cite{43} exists, such that for 
values of $s$ along the backbone of the chain corresponding to distances 
$r(s) > \xi_T$ one expects that excluded volume effects 
are visible and hence Eq.~(\ref{eq2}) should hold. For semidilute 
solutions~\cite{43}, on the other hand, in the good solvent regime 
the inverse effect occurs: there exists a screening length $\xi(c)$ 
depending on the polymer concentration $c$ (also called size of 
``concentration blobs''~\cite{43}), such that excluded volume effects 
are pronounced for $r(s) < \xi (c)$ but are absent for 
$r(s) \gg \xi(c)$. Then Eq.~(\ref{eq3}) holds for the latter 
case and Eq.~(\ref{eq2}) for the former, for rather flexible chains. 
If the chains are semiflexible, in favorable cases (e.g., for simple  
chemical architecture of the polymers) we might observe 
Eq.~(\ref{eq1}) for $1 < s < s^*$ where $s^*$ depends on the 
local intrinsic stiffness of the chain, which we wish to 
characterize by $\ell_p$. Then the  question arises whether 
$s^*$ is smaller than any of the other crossover chemical distances 
(due to marginal solvent quality, described by a Flory-Huggins 
parameter $\chi$ with $(1/2 - \chi) \ll 1$~\cite{43,44}, or due to 
nonzero c) or not. The conclusion of this discussion is that the 
behavior of bond orientational correlations 
$\langle \vec{a}_i \cdot \vec{a}_j \rangle$ is subtle, and not always 
suitable to obtain straightforwardly information on the intrinsic 
stiffness of macromolecules; as a further caveat we mention that in 
general it is also not true that this correlation depends on the 
relative distance $s=|i-j|$ only: it matters also, if one of the 
sites is close to a chain end.

Another popular definition is the local persistence length $\ell_p(i)$ 
defined as~\cite{1,2}
\begin{equation} \label{eq5}
\ell_p(i) / \ell_b=\langle \vec{a}_i \cdot \vec{R} \rangle / 
\langle \vec{a}^2_i \rangle \, .
\end{equation}
However, it has been shown by renormalization group methods that in good 
solvents one has, for $N \rightarrow \infty$, 
$\ell_p (i) \propto [i (N-i)]^{2 \nu -1}$, $i \gg 1$, so the behavior 
of $\ell_p(i)$ in the chain interior clearly is unsuitable to conclude 
anything about the local stiffness of a chain under good solvent 
conditions, and this conclusion has been corroborated by 
simulations~\cite{11,17}. Sometimes it has been argued that a better choice 
is to take the correlation between the first bond vector and the 
end-to-end distance, $\ell_p(1)$~\cite{46}. However, since in a 
macromolecule the chemical nature of the end monomer always differs 
from inner monomers, one can never expect that $\ell_p(1)$ precisely 
characterizes the local stiffness of a linear macromolecule in the 
inner parts of a chain. Moreover, since
$\langle R^2 \rangle$ reflects all the crossovers (due to 
``thermal blobs'' etc.),~\cite{43}, as discussed above, it is 
premature to expect that $\ell_p(1)$ stays unaffected from them. 
We also note that for $d=2$ dimensions under good solvent 
conditions it has been shown~\cite{46a} that 
$\ell_p(1) \propto \ln N \rightarrow \infty$ as 
$N \rightarrow \infty$, so in this case $\ell_p(1)$ clearly is not 
a useful measure of the intrinsic stiffness of a chain at all. 
Since Eq.~(\ref{eq5}) is difficult to extract from any experiments, 
and inconvenient for simulation studies due to high sampling effort, 
we shall not discuss Eq.~(\ref{eq5}) further in the present paper.

\begin{figure}
\begin{center}
% for a multi-line caption
%\onelinecaptionstrue
% for a one-line caption
%\includegraphics{}
\includegraphics[scale=0.19,angle=0]{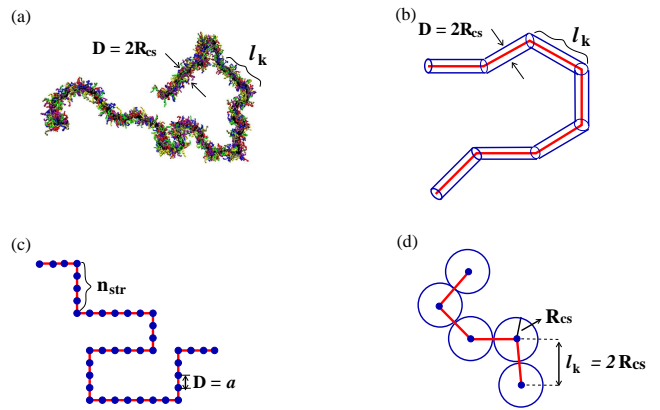}\hspace{0.4cm}
\caption{\label{fig4} Various models of semiflexible polymers,
as discussed in the context of simulations. Case (a)
shows the snapshot picture of a typical conformation of a simulated
bottle-brush polymer using a backbone chain
length $N_b=1027$, side chain length $N_s=24$, projected into the
$xy$-plane (this model is discussed in more detail in Sec. 3).
Case (b) shows a model of freely jointed cylindrical rods of Kuhn
step length $\ell_K$ and diameter $D=2R_{cs}$, with
$R_{cs}$ the cross-sectional radius (if $R_{cs}=0$ this leads to
a simple off-lattice random walk configuration, while excluded
volume interaction is introduced if overlap of the cylinders is
forbidden). Case (c) shows the SAW model on the square lattice
with lattice spacing $a$ ($D=a$ in this case),
where 90$^o$ bends cost an energy $\varepsilon_b \gg k_BT$, so the
chain consists of straight pieces
where $n_{\rm str}$ steps go in the same lattice direction, with
$n_{str} \gg 1$. Case (d) shows a model of tangent hard spheres
with radius $R_{cs}$ (and $\ell_k=2 R_{cs}$).}
\end{center}
\end{figure}

Experimental studies try to extract the persistence length either
from scattering analyses of the single chain structure factor
(e.g.~\cite{28,32,34,35,36,37,38,46b}) or from analyses of
extension versus force measurements of stretched chains
(e.g.~\cite{47,48,49,50,51,52,53,54,55}). However,
the interpretation of the latter experiments must rely on a theoretical
model of the extension versus force curve. While this task is simple
for ideal random walk models of polymers~\cite{4,5,56} and also for
semiflexible polymers when excluded volume is
neglected~\cite{57}, so that the Kratky-Porod (K-P) model~\cite{58} of
worm-like chains can be used, it is very difficult (due to multiple
crossovers~\cite{59,60}) if excluded volume effects are included.
These excluded volume effects cause an intermediate nonlinear
variation of the extension versus force curve (the chain is then a
string of ``Pincus blobs''~\cite{61}), making the estimation of the
persistence length difficult~\cite{60}, and this behavior has also been
verified in recent experiments~\cite{54,55}. Since we have given a
recent extensive discussion of this problem elsewhere~\cite{60}, we shall not
dwell on this problem here further, and focus on the problem how the
persistence length shows up in the single chain structure factor $S(q)$.
Here the key idea is that the scattering intensity $S(q)$ at scattering
wavenumber $q$ yields information on the structure of the macromolecule
at a length scale $\lambda = 2 \pi/q$. This problem also is subtle,
even in the framework of simple models (see Fig.~\ref{fig4}a,b,c,d)
used for simulations. If $\lambda$ is of the scale of the cross sectional
radius $R_{cs}$ for the models (a,b,d) or the lattice spacing in (c),
local structure on the scale of effective subunits is revealed: soft
(a) versus hard (b) effective cylinders, hard spheres in (d),
but one could also conceive a chain where soft spheres are
jointed, etc. When one considers semiflexible chains with no
excluded volume, the persistence length $\ell_p$ would be just
one half of the step length $\ell_K$ in cases (b), (d), where one
then requires a strong bond angle potential to make these chains
semiflexible rather than flexible; however, as emphasized above,
such models neglecting excluded volume completely will inevitably imply
Eq.~(\ref{eq1}), which is inappropriate for real polymers under
all physically possible conditions. So the information on chain
stiffness, as described by the persistence length, is hidden in some
intermediate range of wavenumbers. E.g., for the model (c), which will
be used extensively in the rest of the paper (but in $d=3$ dimensions,
since the case of $d=2$ is rather special~\cite{60} as will be
discussed below), we need wavenumbers in the range
$2 \pi / \sqrt{\langle R^2_g \rangle} \ll q \ll 2 \pi/a$, where $a$
is the lattice spacing. The aim of the present paper is to present
a discussion of how one can obtain detailed information on intrinsic chain
stiffness from the gyration radius of the macromolecules
and from the structure factor $S(q)$ in the suitable intermediate range of
wavenumbers $q$.

The outline of our paper is as follows: in the next section, 
we summarize some pertinent theoretical results on $S(q)$. 
In the third section, our Monte Carlo simulation methods are briefly 
described. In the fourth section, a comparative discussion of simulation 
results for two models is given, the bond fluctuation model of 
bottle-brushes (c.f.~Fig.~\ref{fig4}a), 
and the self-avoiding walk model on the simple cubic lattice 
with variable bending energy (cf. Fig.~\ref{fig4}c). The final section 
contains our conclusions.

\section{SOME THEORETICAL RESULTS ON THE STRUCTURE FACTOR OF 
ISOLATED MACROMOLECULES IN SOLUTION}

We consider here a single macromolecule with linear chain architecture, 
assuming a sequence of $N+1$ (effective) monomeric units at positions 
$\vec{r}_j$, $j=1,2, \cdots, N+1$, with effective bond vectors 
$\vec{a}_j=\vec{r}_{j+1} - \vec{r}_j$, $j=1, \cdots, N$. We have in mind 
application to standard polymers like polystyrene (disregarding here the 
scattering from the side groups that are attached to the backbone of the chain, 
see e.g. Rawiso et al.~\cite{46b} for a discussion of this problem in an 
experimental context). We also have in mind application to the scattering 
from the backbone of bottle-brush polymers (this is experimentally directly 
accessible from neutron scattering~\cite{28} if selective deuteration only 
of the backbone is used, while in the case of deuteration of the whole 
macromolecules~\cite{32,34} this information can be inferred only indirectly). 
Due to the restriction to ``effective monomeric units'' rather than talking 
about the scattering from individual atoms with the appropriate scattering 
lengths, we clearly disregard information on the scale of the length of an 
effective bond, but we then need not discuss experimental problems such as 
contrast factors between the scattering from the macromolecule and the 
solvent~\cite{46b}. The effect of the cross-sectional structure of the 
chain (finite chain thickness $D$) is not explicitly considered as well 
(experimentally this problem often is approximated in terms of the 
Guinier~\cite{71} approximation, writing the observed scattering 
intensity $S_{\rm obs}(q)=S(q) \exp (- q^2 R^2_c/2)$, with $R_c$ some 
``effective'' cross-sectional radius of the chain~\cite{46b}. Thus only 
wavenumbers $q D\ll 2 \pi$ are physically meaningful: in the case of the 
lattice model, Fig.~\ref{fig4}c, $D=a$, of course. The structure factor 
then is defined as
\begin{equation} \label{eq6}
S(q)=\frac{1}{(N+1)^2} \Big\langle \sum\limits_{j=1}^{N+1} \, 
\sum\limits_{k=1}^{N+1} \exp \Big[i\vec{q} \cdot 
(\vec{r}_j - \vec{r}_k)\Big] \Big\rangle \, ,
\end{equation}
and does not depend on the direction of the scattering 
wavevector $\vec{q}$. In $d=3$ dimensions, it has the small $q$ expansion
\begin{equation} \label{eq7}
S(q) = 1 - \langle R^2_g \rangle q^2 / 3 + \cdots \, , \quad q \rightarrow 0\,,
\end{equation}
where the mean square gyration radius $\langle R^2_g \rangle$ enters
\begin{equation} \label{eq8}
\langle R^2_g \rangle = \frac{1}{(N+1)^2} \Big\langle 
\sum\limits_{j=1}^{N+1} \, \sum\limits_{k=j+1}^{N+1}
(\vec{r}_j - \vec{r}_k)^2 \Big\rangle \, .
\end{equation}
Other characteristic lengths of the chain molecule are the mean square end-to-end distance
\begin{equation} \label{eq9}
\langle R^2 \rangle = \Big\langle \Big(
\sum\limits_{i=1}^N \vec{a}_i\Big)^2 \Big\rangle
\end{equation}
and the contour length
\begin{equation} \label{eq10}
L=N \ell_b \, ,
\end{equation}
but neither of these lengths can be inferred directly from the scattering. 
For chains in dense melts or in dilute solutions under Theta conditions 
one typically uses an ideal chain approximation 
(disregarding, e.g., logarithmic corrections at the 
Theta point~\cite{3,43,62})
\begin{equation} \label{eq11}
\langle R^2 \rangle = 6 \langle R^2_g \rangle = C_\infty \ell^2_b N \, , 
\quad N \rightarrow \infty \, ,
\end{equation}
with $C_\infty$ a characteristic constant~\cite{1,2,3,4,5}. 
In this case one introduces an equivalent freely jointed Kuhn chain with 
the same contour length, $\langle R^2 \rangle= n_k \ell^2_K$, 
where $n_K$ is the number of equivalent Kuhn segments and $\ell_K$ 
their length,
\begin{equation} \label{eq12}
\ell_K=C_\infty \ell_b \, , \quad n_K=N/C_\infty \, , \quad 
N \rightarrow \infty \, .
\end{equation}
For a semiflexible worm-like chain with $C_\infty \gg 1$ Eq.~(1) holds.
However, since under Theta conditions (and melts) 
Eq.~(\ref{eq12}) is approximately true, one finds
\begin{equation} \label{eq13}
\ell_p= 3 \langle R^2_g \rangle /(N \ell_b) \, , \quad 
\langle R^2 \rangle = 2 \ell_p \ell_b N \, , \quad N \rightarrow \infty \, ,
\end{equation}
if the relation $\ell_p=\ell_k/2= C_\infty \ell_b/2$ then simply is taken 
as an alternative definition of a persistence length. For the simple SAW 
model of Fig.~\ref{fig1} this gives $\ell_p=0.94$ lattice spacings: 
but as expected, using this value in the simple exponential 
$\exp(-s \ell_b/\ell_p)$ one does not obtain a description of the actual 
data in Fig.~\ref{fig1} on the basis of this description, because the 
actual behavior of bond orientational correlations is a power law decay, 
Eq.~(\ref{eq3}). Note, however, that the relation $\ell_p=\ell_K/2$ 
makes only sense for semiflexible chains for which $C_\infty \gg 1$ at 
the Theta point, which is not the case for the model of Fig.~\ref{fig1}.

In the case of good solvent conditions excluded volume interactions 
invalidate Eq.~(\ref{eq11}) and one finds instead~\cite{3,7,43,62}
\begin{equation} \label{eq14}
\langle R^2 \rangle =2 \ell^R_p \ell_b N^{2 \nu} \, , 
\quad \langle R^2_g \rangle =\frac{1}{3} \ell^{R_g}_p \ell_b N^{2 \nu} \, , 
\quad N \rightarrow \infty
\end{equation}
with~\cite{7} $\nu \approx 0.588$ instead of the mean field value 
$\nu_{MF}=1/2$ that appears in Eq.~(\ref{eq13}). Note that we have 
defined the prefactors of the relations $\langle R^2 \rangle \propto N^{2\nu}$,
$\langle R^2_g \rangle \propto N^{2 \nu}$ in Eq.~(\ref{eq14}) in complete 
analogy with Eq.~(\ref{eq13})~\cite{11}, but we shall see shortly that 
the lengths $\ell^R_p$, $\ell^{R_g}_p$ do not play the role of a 
persistence length that describes the local intrinsic stiffness of the chains.

For a better understanding of this problem, in particular when $\ell_p$ is 
very large, it is of interest to consider the crossover from the rod limit 
(that occurs for $L < \ell_K$, i.e. $n_K <1$) to the Gaussian coil limit. 
This problem can be worked out easily for various models of discrete 
chains~\cite{1,2,3,4,5} as well as for the Kratky-Porod model. 
Describing the chain by a continuous curve $\vec{r}(s)$, $s$ being 
the curvilinear coordinate along the chain contour, 
the potential energy of a particular conformation of the chain is given by
\begin{equation} \label{eq15}
\mathcal{H} = \frac{\kappa}{2} \int\limits_0^L \Big(
\frac{\partial^2 \vec{r} (s)}{\partial s^2}\Big)^2 \, ds\,, 
\quad \kappa=k_BT \ell_p \, (d=3) \, .
\end{equation}
In Eq.~(\ref{eq15}) it is clearly assumed that $\kappa$ is a constant, 
independent of the contour length $L$ (or chain length $N$, respectively), 
and the same holds for $\ell_p$. The physical interpretation of $\kappa$ is 
in terms of the local bending stiffness of the chain. Formula~(\ref{eq15}) 
can be used for arbitrary values of the ratio $L/\ell_p=n_p$, and one can 
show~\cite{58,63}
\begin{equation} \label{eq16}
\frac{\langle R^2 \rangle}{2 \ell_pL} =1-
\frac{1}{n_p} \Big[1-\exp (-n_p)\Big] \, ,
\end{equation}
and 
\begin{equation} \label{eq17}
\frac{3 \langle R^2_g\rangle}{\ell_pL} = 1 - \frac{3}{n_p} + 
\frac{6}{n^2_p}-\frac{6}{n^3_p} \Big[1 -\exp (-n_p)\Big] \, .
\end{equation}
One immediately recognizes that for $n_p=2n_K \rightarrow \infty$ one 
recovers Eq.~(\ref{eq13}), while in the opposite limit the results 
for rigid rods of length $L$ are obtained,
\begin{equation} \label{eq18}
\langle R^2\rangle =12 \langle R^2_g \rangle =L^2 \,, \quad n_p \ll 1 \, .
\end{equation}
However, the generalization of these results to the good solvent case,
where excluded volume matters, is not straightforward. 
Of course, for $n_p \ll 1$ 
excluded volume is irrelevant, Eq.~(\ref{eq18}) remains valid. 
It turns out, however, that in $d=2$ Eqs.~(\ref{eq16}), (\ref{eq17}) 
are not valid at all, one has no regime of Gaussian chain behavior as 
described in Eq.~(\ref{eq13}), and rather near $n_p=1$ a crossover 
from rigid rod behavior 
to the behavior of two-dimensional self-avoiding walks 
occurs~\cite{60,64} ($\nu=3/4$)
\begin{equation} \label{eq19}
 \langle R^2 \rangle \propto \langle R^2_g \rangle \propto
\ell^{1/2}_p L^{3/2} \, , \quad L > \ell_p.
 \end{equation}
For $d=3$, however, Eqs.~(\ref{eq16}), (\ref{eq17}) for semiflexible 
chains remain valid for $n_p <n^*_p (\ell_p)$ where 
$n^*_p (\ell_p \rightarrow \infty) \rightarrow  \infty$. 
This crossover contour length $L^*=n^*_p \ell_p$ has first been 
estimated by a Flory argument as~\cite{60,65,66}
\begin{equation} \label{eq20}
L^* \propto \ell^3_p/D^2 \, , \quad n^*_p \propto(\ell_p/D)^2 \, .
\end{equation}
Note, however, that Flory arguments imply $\nu=3/5$ in $d=3$ (rather than 
the precise value $\nu \approx 0.588$~\cite{7}) and cannot predict any 
prefactors in Eq.~(\ref{eq20}); they are based on a crude balancing 
of the elastic energy of chain stretching (taken as Gaussian) and a mean 
field estimate of binary interactions: having in mind a model description 
as in Fig.~\ref{fig4}b, one takes the second virial coefficient 
proportional to the rod volume on the scale of the persistence length, 
$\upsilon_2 \propto\ell^2_p D$, and in this way the effective chain 
diameter $D$ enters the estimated Eq.~(\ref{eq20})~\cite{60,65,66}. 
Numerical results, however, seem to suggest that rather~\cite{60} 
$n^*_p \propto(\ell_p/D)^\zeta$ with an exponent $\zeta \approx 1.5.$

In any case, the conclusion of this discussion is that for semiflexible 
chains in $d=3$ the mean square radii as a function of the reduced contour 
length $n_p=L/\ell_p$ exhibit two successive crossovers, from rods to Gaussian 
coils near $n_p=1$ and from Gaussian coils to swollen chains (described by 
Eq.~(\ref{eq14})) near $n_p=n^*_p$. These two crossovers have in fact 
been seen nicely in both experiment~\cite{67} and computer simulation of 
the model of Fig.~\ref{fig4}c~\cite{60,68,69}.

We now turn to a discussion how these behaviors show up in the scattering 
function $S(q)$ at larger wavenumbers, when Eq.~(\ref{eq7}) does not 
hold. In the regime $n_p < 1$, when the chain behaves like a rigid rod, 
one can work out the scattering function in the continuum limit as~\cite{70}

\begin{equation} \label{eq21}
S_{\rm rod} (q) = \frac{2}{qL} \Big[\int\limits^{qL}_0 \, 
dx \frac{\sin x}{x} -\frac{1-\cos (qL)}{qL} \Big]
\end{equation}
while for a discrete chain of $N+1$ scatterers along a rod of length 
$L=N \ell_b$ one has
\begin{eqnarray} \label{eq22}
S_{\rm rod} (q) 
= \frac{1}{N+1} \Big[-1 + \frac{2}{N+1} \, 
\sum\limits_{k=0}^N \, &&(N+ 1 - k) \frac{\sin (q\ell_b k)}{q \ell_b k}\Big],
\nonumber \\ 
&&q \ell_b < 2 \pi \, .
\end{eqnarray}
It is noteworthy to recall that the large $q$-limit of 
Eq.~(\ref{eq21}) contains information on the contour length $L$ 
and shows a $1/q$ decay,
\begin{equation} \label{eq23}
S_{\rm rod} (q \rightarrow  \infty) = \pi / (q L) \, .
\end{equation}
In the Gaussian regime, that applies for chain lengths that correspond 
to $1 \ll n_p <n^*_p (\ell_p)$ in $d=3$, the structure factor $S(q)$ 
is described by the well-known Debye function,
\begin{equation} \label{eq24}
S_{\rm Debye} (q) = \frac{2}{X} \Big\{ 1- \frac{1}{X} 
\Big[1-\exp (-X)\Big]\Big\} \,, \quad X \equiv 
q^2 \langle R^2_g\rangle \, .
\end{equation}
For small $X$, Eq.~(\ref{eq24}) reduces to Eq.~(\ref{eq7}), 
as it must be, while for large $X$ Eq.~(\ref{eq24}) yields 
$S_{\rm Debye}(q) \approx 2/X=2/[q^2 \langle R^2_g \rangle ]$. While 
for flexible chains at the Theta point 
Eq.~(\ref{eq24}) is expected to hold for large $q$, up to $q \ell_b$ 
of order unity where effects due to the local structure of monomeric 
units comes into play, the validity of Eq.~(\ref{eq24}) for 
semiflexible chains is much more restricted, since then the rod to coil 
crossover matters also with respect to the intrinsic structure of these 
polymers, as it is probed by $S(q)$. In oder to discuss this problem, 
it is useful to cast $S(q)$ in the representation of the so-called 
Kratky plot~\cite{71}, $q LS(q)$ is plotted as a function of $qL=Y$. 
For rigid rods, one simply would have a linear
increase of $qLS(q)$ with $Y$ for small $Y$, which smoothly crosses 
over near $Y=1$ to a flat plateau (which has the value $\pi$, 
cf. Eq.~(\ref{eq23})). For chains where intermonomer distances 
follow Gaussian distributions, at all scales, the Kratky plot exhibits 
a maximum at $Y_{\rm max}$, and then a crossover to a decay proportional 
to $Y^{-1}$ occurs. To locate this maximum, it is convenient to write 
$qLS(q)$ as \,$ \sqrt{X} (L/\sqrt{\langle R^2_g \rangle}) S_{\rm Debye}(q)$ 
as a function of $X$, noting that the maximum occurs at 
$X_{\rm max} \approx 2.13$, i.e. the Kratky plot has its maximum at 
$Y_{\rm max} \approx \sqrt{2.13} L/ \sqrt{\langle R^2_g\rangle}$, and the 
height of this maximum also is of order $L/\sqrt{\langle R^2_g \rangle}$. 
Using now Eq.~(\ref{eq13}) in the form $\langle R^2_g \rangle= L \ell_p/3$, 
we recognize that the maximum of the Kratky plot occurs at
\begin{equation} \label{eq25}
(qL)_{\rm max} \approx \sqrt{6.4} (L/\ell_p)^{1/2} \, , 
\quad 1 \ll L/\ell_p < n_p^*(\ell_p)
\end{equation}
and also the height of this maximum scales proportional to 
$\sqrt{L/\ell_p}$. However, while for flexible chains under 
Theta conditions (for which $\ell_p$ and $\ell_b$ are of the 
same order), one observes on the
Kratky plot for $Y >> Y_{\rm max}$ a decay $q LS(q) \propto q^{-1}$, 
for semiflexible chains a crossover from this decay to the plateau 
value $\pi$ (given by 
Eq.~(\ref{eq23})) is expected. This is also true for semiflexible 
chains under good solvent conditions, if the persistence length $\ell_p$ 
is large enough so that $n_p(=L/\ell_p) < n^*_p (\ell_p)$, and hence 
excluded volume effects still can be ignored. The description of this 
decay of the structure factor from its peak towards this so-called 
``Holtzer plateau''~\cite{72} has been a longstanding problem in the literature~\cite{73,74,75,76,77,78,79,80,81,82,83,84,85,86,87,88,89,90,91,92,93,94}. 
Only in the limit $N \rightarrow \infty$ a simple explicit result 
derived from Eq.~(\ref{eq15}) is available~\cite{78},
\begin{equation} \label{eq26}
LqS(q) = \pi + \frac{2}{3} \Big(q \ell_p\Big)^{-1} \, , 
\quad L \rightarrow \infty\, ,  \quad q \rightarrow \infty \, ,
\end{equation}
but we should keep in mind that the limit $q \rightarrow \infty$ is 
well-defined for a simple mathematical continuum model such as 
Eq.~(\ref{eq15}), while for real chains (and for simulations) the 
regime $q \ell_b > 2 \pi$ is not at all meaningful. Although the decay 
$S(q) \propto q^{-2}$, that Eq.~(\ref{eq26}) predicts for 
$q \ell_p \ll 1$, is compatible with the power law decay of the 
Debye function, Eq.~(\ref{eq24}), for large $q$,
\begin{equation} \label{eq27}
S(q) \approx  \frac{2}{q^2 \langle R_g^2\rangle} 
\rightarrow q L S(q) \approx 6 (q \ell_p)^{-1}
\, , \quad q \rightarrow \infty \,,
\end{equation}
the prefactor in Eq.~(\ref{eq27}) is by a factor of $9$ larger 
than the prefactor of the $q^{-1}$ term in Eq.~(\ref{eq26}), so 
both Eqs.~(\ref{eq24}) and (\ref{eq26}) are inconsistent 
with each other. This inconsistency is due to the fact that 
Eq.~(\ref{eq26}) is only accurate for $q \ell_p > 3$, it should not 
be used for small $q \ell_p$. 
After many less successful attempts, Kholodenko~\cite{84,85,86,87} achieved 
a description which interpolates between the limiting cases of rigid rods 
and of Gaussian coils, capturing the scattering law of both limits exactly,
but deviating from the exact result (``exact'' refers to the Hamiltonian 
Eq.~(\ref{eq15}), so no excluded volume effects are being accounted for) 
in the intermediate regime; this exact behavior is known from systematic expansions~\cite{92,93,94} whose use requires heavy numerical work, and will 
not be considered here. Recently we have shown~\cite{95} that the exact 
method of Stepanow~\cite{93,94} deviates only very little from the 
approximation of Kholodenko~\cite{87}, which can be cast in the form
\begin{equation} \label{eq28}
S(q) =\frac{2}{x} \Big[I_1 (x) - \frac{1}{x} I_2 (x)\Big]\, , 
\quad x= 3L/ 2 \ell_p \, ,
\end{equation}
where
\begin{eqnarray} \label{eq29}
I_n(x) & = &\int\limits^x_0 d z \, z^{n-1} f(z) \,, \nonumber \\
f(z)&=&
\begin{cases}
\frac{1}{E} \frac{\sinh (Ez)}{\sinh z} \, , \quad q \leq 3 / 2 \ell_p \,,\\
\frac{1}{E'} \frac{\sin (E' z)}{\sinh z}
\,, \quad q > 3 / 2 \ell_p \,, 
\end{cases}
\end{eqnarray}
with
\begin{equation} \label{eq30}
E= [1-(2 q \ell_p/3)^2 ]^{1/2} \, , \quad 
E'=[(2 q \ell_p/3)^2 -1]^{1/2} \, .
\end{equation}
We stress that all these analytical results 
Eqs.~(\ref{eq11})-(\ref{eq13}), (\ref{eq16}), (\ref{eq17}),
(\ref{eq24})-(\ref{eq30}) are only applicable if excluded volume 
effects are negligible. When we consider very long semiflexible chains, 
such that $n_p=L/\ell_p > n^*_p (\ell_p)$, we expect that the Gaussian 
results $\langle R^2 \rangle=2 \ell_pL= 2 \ell^2_p n_p$ and 
$\langle R^2_g \rangle =(1/3) \ell_p L =(1/3) \ell^2_p \, n_p$ hold roughly
up to $n^*_p (\ell_p)$, and there a smooth crossover to the excluded 
volume power laws, Eq.~(\ref{eq14}), occurs. We first note that 
hence $n^*_p (\ell_p)$ corresponds to a crossover radius $R^*$ of the 
chains as well, $R^{*2}= 2 \ell_p L^*=2 \ell^2_p n^*_p$. Omitting factors 
of order unity, we conclude
\begin{equation} \label{eq31}
R^* = \sqrt{\ell_p L^*}= \ell_p \sqrt{n^*} \propto \ell^2_p/D \quad,
\end{equation}
where in the last step Eq.~(\ref{eq20}) was used. For 
$n_p > n^*_p (\ell_p)$ we hence expect, invoking the fact that the crossover 
in the linear dimensions for $n_p=n^*_p$ should be smooth, 
\begin{eqnarray} \label{eq32}
\langle R^2 \rangle &=&R^{*2} (n_p/n^*_p)^{2 \nu}  \nonumber\\
&\propto& \ell^4_p / D^2 (D/\ell_p)^{4 \nu} \, n^{2 \nu}_p 
\approx\ell^2_p \Big(\frac{D}{\ell_p} \Big)^{2/5} n_p^{6/5} \, ,
\end{eqnarray}
where in the last step the Flory estimate $\nu \approx 3/5 $ was 
used (recall that in Eq.~(\ref{eq31}) the exponent $\zeta$ 
defined above has also been put to its Flory value, $\zeta=2)$. In terms 
of $N$ and $\ell_p$, Eq.~(\ref{eq32}) becomes 
$\langle R^2 \rangle \propto \ell^{6/5}_b (\ell_pD)^{2/5} N^{6/5}$. 
In terms of the constant $\ell^R_p$ defined in Eq.~(\ref{eq14}), 
we would have $\ell_p^R \propto \ell_b^{1/5} (\ell_p D)^{2/5}$.

\begin{figure*}
\begin{center}
% for a multi-line caption
%\onelinecaptionstrue
% for a one-line caption
%\includegraphics{}
\includegraphics[scale=0.30,angle=0]{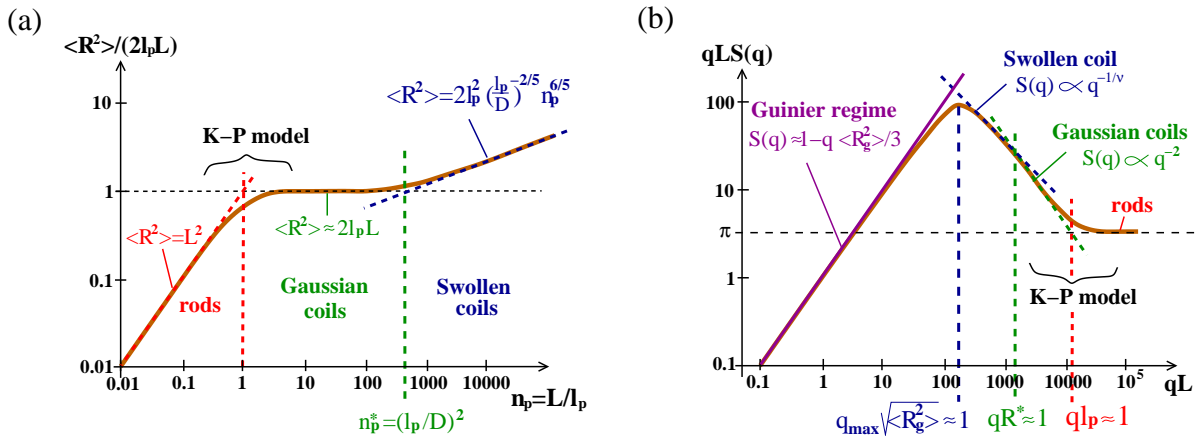}\hspace{0.4cm}
\caption{\label{fig5} (a) Schematic plot of the normalized mean square
radius $\langle R^2 \rangle/(2 \ell_p L)$ versus $n_p=L/\ell_p$
(apart from a factor of 2 this is the number of Kuhn segments),
on log-log scales. The Kratky-Porod (K-P) model describes the crossover
from rods $(\langle R^2 \rangle =L^2)$ to Gaussian coils
$(\langle R^2 \rangle =2 \ell_p L)$. At $n^*_p=(\ell_p/D)^2$,
according to the Flory theory a crossover to swollen coils occurs,
where $\langle R^2 \rangle \propto n^{2 \nu}_p$ with $\nu= 3/5$
(according to the Flory theory). (b) Schematic Kratky plot of the
structure factor of a semiflexible polymer, $q LS(q)$ plotted vs.~$qL$,
on log-log scales. Four regimes occur: in the Guinier-regime,
$S(q) \approx 1-q^2 \langle R^2 \rangle/3$; it ends at the maximum
of the Kratky plot, which occurs roughly at
$q_{\rm max} \sqrt{\langle R^2 \rangle} \approx 1$
(constants of order unity being ignored throughout).
For very large $L$ then a regime of swollen coils with
$S(q) \propto q ^{-1/\nu}$ is observed, until near $q R^* \approx 1$
a crossover to Gaussian coil behavior occurs ($R^* \approx \ell^2_p/D$).
In the Gaussian coil regime $S(q) \propto q^{-2}$, until at
$q \ell_p$ of order unity the crossover to the rod-like regime
occurs ($qLS (q)= \pi)$. Only the latter two regimes are captured by the Kratky-Porod model.}
\end{center}
\end{figure*}

The consequences for the scattering function $S(q)$ are now clear, 
since the gyration radius shows the same scaling behavior as 
$\langle R^2 \rangle$, apart from prefactors of order unity. 
Hence we have
\begin{equation} \label{eq33}
\sqrt{\langle R^2_g \rangle} \propto \ell_b^{3/5} (\ell_p D)^{1/5} \, N^{3/5}
\end{equation}
and only for $1 / \sqrt{\langle R^2_g \rangle} < q < 1/R^*$ we can  
expect to see the nontrivial power law
\begin{equation} \label{34}
S(q) \propto q^{-1/\nu} \, ,
\end{equation}
while at $q^*$ defined from $q^*R^*=1$ we have a smooth crossover to the 
standard Debye law, $S(q) \propto q^{-2}$. Near $q \ell_p=1$ then a 
smooth crossover to the rod-like scattering law $S(q) \propto q^{-1}$ 
occurs. So the three power laws for the radii as a function of chain 
length $(\sqrt{\langle R ^2\rangle} \propto N$ in the rod regime,
$\propto N^{1/2}$ in the regime of Gaussian coils, and 
$\propto N^\nu$ in the regime of swollen coils) find their counterpart 
in the scattering function, if $N$ is large enough. The schematic 
Fig.~\ref{fig5} illustrates these crossover behaviors. The three 
regimes of the $\langle R ^2 \rangle$ versus $N$ (or $n_p$, respectively) 
curve, namely rods, Gaussian coils, and swollen coils
(Fig.~\ref{fig5}a) appear in the $S(q)$ vs.~$q$ curve (or $q LS(s)$ 
vs.~$qL$-curve, in the Kratky representation) in inverse order: the rods 
occur for large $q$, then occurs a first crossover to Gaussian coils, 
and a second 
crossover to swollen coils. Of course, if the chains are very stiff but not 
extremely long, it may be that the regime $n_p > n^*_p$ is not reached: 
then in part (a) the swollen coil regime is absent, and in part (b) as well: 
then the K-P model can describe $S(q)$ fully, including the regime of the 
maximum of the Kratky plot. Since the crossovers are smooth, it may be 
difficult to identify the different power laws in Fig.~\ref{fig5}b 
in practice, however.

We also note that the different regimes are also only well separated if 
both $\ell_p$ is very large (in comparison to $\ell_b$) and also 
$\ell_p/D$ needs to be very large. If $\ell_p$ is very large, but $D$ 
also (as in the case of bottle-brush polymers~\cite{11,17,68,69}) then 
the regime of Gaussian coils disappears from both Fig.~\ref{fig5}a 
and ~\ref{fig5}b, and the K-P model loses its applicability.

\section{MONTE CARLO SIMULATION METHODS AND MODELS}

In the present work, we focus on lattice models exclusively, because 
for them particularly efficient simulation methods exist; pertinent 
work on coarse-grained off lattice models of bottle-brush polymers 
studied in Molecular Dynamics methods for variable solvent 
quality~\cite{96} will be mentioned in the conclusions section.

The archetypical lattice model of a polymer is the self-avoiding 
walk on the simple cubic lattice~\cite{97}. Each effective monomer 
takes a single lattice site, the length of an effective bond is the 
lattice spacing, so adjacent monomers along the chain are nearest 
neighbors on the lattice. Double occupancy of lattice sites being 
forbidden, excluded volume interactions under very good solvent 
conditions are modelled.

The properties of this basic model are very well established~\cite{98}. 
Solvent quality can be included as a variable into this model implicitly, 
by allowing for an (attractive) energy $\varepsilon$ that is won if two 
monomers (that are not nearest neighbors along the chemical sequence of 
the chain) are nearest neighbors on the lattice. One then finds that the 
Theta point, at which (apart from logarithmic corrections~\cite{3,62}) the 
mean square radius $\langle R^2 \rangle $ scales like a Gaussian chain, 
$\langle R^2 \rangle \propto N$, occurs for 
$q \equiv \exp (\varepsilon/k_B T)=q_\theta\, (\approx 1.3087)$~\cite{12}. 
On the other hand, if one introduces an energy cost $\varepsilon_b$ 
whenever the walk makes a turn by $\pm 90^o$ (of course, reversals by 
180$^o$ are forbidden, because of the excluded volume constraint), 
one can vary the local intrinsic stiffness of the chain 
(cf. Fig.~\ref{fig4}c, which illustrates this model for $d=2$ dimensions). 
For $q_b=\exp (-\varepsilon_b/k_BT)=1$ one recovers the standard SAW, 
while the limit $\varepsilon_b \rightarrow \infty$ corresponds to rigid straight rods. 
Following up on our previous work~\cite{11,60,68,69,95}, we shall focus on 
this model in the present paper, applying the pruned-enriched Rosenbluth 
method (PERM)~\cite{12,99}. PERM is a biased chain growth algorithm with 
resampling and allows to get accurate data up to $N=50000$ for this 
model~\cite{68,69}. PERM yields a direct estimate of the partition function 
of a self-avoiding walk with $N$ steps and $N_{\rm bend}$ $90^o$-bends
\begin{equation} \label{eq35}
Z_N (q_b) = \sum\limits_{\rm config.} C_{N, N_{\rm bend}} \, q_b^{N_{\rm bend}}
\end{equation}
where $C_{N, N_{\rm bend}}$ is the number of configurations of SAW's 
with $N$ bonds and a number $N_{\rm bend}$ of $\pm 90^o$ turns. 
It would be interesting 
to extend the approach from athermal semiflexible chains 
$(q=\exp (\varepsilon/k_BT)=1)$ to semiflexible chains in solvents of 
variable quality $(q > 1)$, which would mean an estimation of
\begin{equation} \label{eq36}
Z_N (q,q_b)= \sum\limits_{\rm config.} \, C_{N, N_{\rm bend}, N_{\rm pair}} 
q_b^{N_{\rm bend}} q^{N_{\rm pair}} \, ,
\end{equation}
with $N_{\rm pair}$ the number of nonbonded nearest neighbor pairs 
of monomers in the considered configuration. However, we are not aware 
of any study of the full problem, Eq.~(\ref{eq36}), yet.

Sampling suitable data on the monomer coordinates of the configurations 
that contribute to the partition function Eq.~(\ref{eq35}), one can 
obtain reasonably accurate estimates of the radii and of $S(q)$, as 
defined in Sec. 2.

The second model that is studied here is the bond fluctuation model of 
bottle-brush polymers. In the bond fluctuation model~\cite{100,101,102}, 
each effective monomer blocks all eight corners of the elementary cube 
of the simple cubic lattice from further occupancy. Two successive 
monomers along a chain are connected by a bond vector $\vec{\ell}_b$, 
chosen from the set 
$\{ (\pm 2,0,0)$,  $(\pm 2, \pm 1, 0)$, $(\pm 2, \pm 1, \pm 1)$, 
$(\pm2, \pm 2, \pm 1)$, 
$(\pm 3,0,0)$, $(\pm 3, \pm 1,0)\}$, including also all permutations. 
Originally configurations were relaxed by an algorithm where a monomer 
of the chain is chosen at random, and one also randomly chooses one of the six 
directions ($\pm x$, $\pm y$, or $\pm z$), respectively, and attempts 
to move the monomers by one lattice unit in the chosen direction. 
Of course, the move is accepted only if it does not violate excluded 
volume or bond length constraints. This move is called the ``L6''
move. Recently Wittmer et al.~\cite{9} provided evidence that a 
much faster algorithm results if one allows monomers to move 
to one of the 26 nearest and next nearest neighbor 
sites surrounding a monomer. With this ``L26'' move bonds can 
cross one another, and while such moves do not correspond to a 
real dynamics of macromolecules, it leads to a much faster 
exploration of phase space and hence a faster equilibration~\cite{103}.

This model for linear polymers is generalized to the bottle-brush architecture 
by adding side chains at regular spacings $1/\sigma$ (which must be integer, 
e.g. for $\sigma=1/2$ a side chain is attached to every second monomer of 
the backbone; the densest packing that is studied here is $\sigma=1)$. 
The side chains have chain length $N_s$, and are described by the bond 
fluctuation model as well. Furthermore, one more monomer is added to each 
chain end, to clearly identify the latter. The number $N_b$ of monomers 
that constitute the backbone then is related to the number of side chains 
$n_c$ via $N_b=(n_c-1) / \sigma +3$ and the total number of monomers of 
the bottle-brush polymer is $N_{\rm tot}=N_b + n_c N_s$. For the sake of 
computational efficiency, the L26 move is combined with Pivot
moves~\cite{98}. We refer to~\cite{17,103} for implementation details.

\begin{figure}
\begin{center}
% for a multi-line caption
%\onelinecaptionstrue
% for a one-line caption
%\includegraphics{}
\includegraphics[scale=0.40,angle=0]{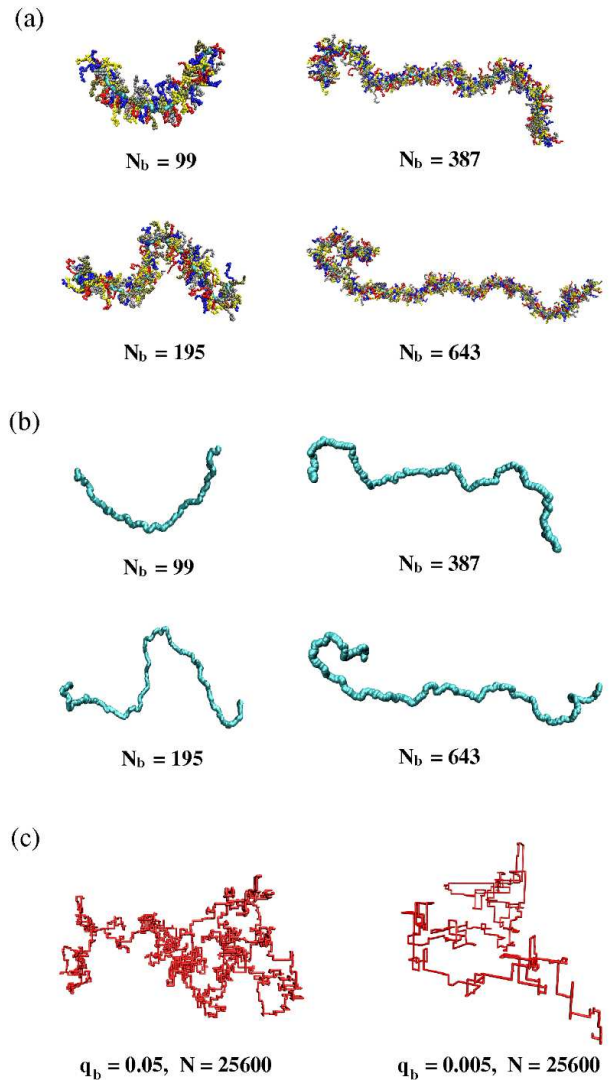}
\caption{\label{fig6} (a) Snapshot pictures of bottle-brush polymers
as described by the bond-fluctuation model, for side chain length
$N_s=24$, and backbone chain length $N_b=99$, $195$, $387$, and $643$. (b)
Same as (a), but displaying the backbone of these bottle-brush
polymers (c) Snapshot pictures of the SAW model with $N=25600$ and
two choices of $q_b$, $q_b=0.05$ and $0.005$.}
\end{center}
\end{figure}

As an example for well-equilibrated bottle-brush polymers as studied 
in~\cite{11,17,68,69} and in the present paper, Fig.~\ref{fig6}a shows 
selected snapshot pictures for side chain length $N_s=24$ and various 
backbone chain lengths $N_b$. According to the visual impression, 
it seems rather natural to describe these bottle-brush polymers by the 
worm-like chain model, but as we shall see below, this conclusion would 
be totally misleading. Experimentalists often are led to a similar 
conclusion from microscope images of semiflexible polymers (e.g. DNA) 
adsorbed at a substrate (see e.g.~\cite{104,105,106}). However, such a 
conclusion is misleading for several reasons: (i) depending on the speed 
of adsorption of the polymer on the substrate, the conformation of the 
adsorbed polymer may be a frozen ``projection'' of the three-dimensional 
coil, which did not have enough time to relax to the two-dimensional 
equilibrium. (ii) In $d=2$ dimensions, excluded volume forces render 
the Kratky-Porod (K-P) model of worm-like chains 
inapplicable~\cite{60,64}, one encounters a direct crossover from the 
rod regime to two-dimensional self-avoiding walk behavior 
(cf. Eq.~(\ref{eq19})) when the contour length $L$ exceeds the 
persistence length $\ell_p$. One also should note that the persistence 
length of a polymer in $d=2$ dimensions is not at all identical to the 
persistence length of the same polymer in $d=3$ dimensions~\cite{60,64}. 
The experimental work (see, e.g.,~\cite{104,105,106})
seems to be unaware of these 
problems and the resulting conclusions from this work need to be 
considered with care. It is also interesting to note that the snapshot 
pictures of the semiflexible SAW model (Fig.~\ref{fig6}c) do not yield 
an immediate visual impression that the chains can be described by the 
K-P model, because of the 90$^o$ kinks; however, as we shall see, despite 
this difference of the local structure the statistical properties on the 
mesoscopic length scales are well described by the K-P model, for 
$q_b \leq10^{-2}$ and $n_p$ less than $n^*_p(\ell_p)$. Thus, we argue 
that on the basis of the inspection of AFM images of semiflexible polymers 
one should be very careful on drawing conclusions which model is appropriate 
to describe these polymers.

\begin{figure}
\begin{center}
% for a multi-line caption
%\onelinecaptionstrue
% for a one-line caption
%\includegraphics{}
(a)\includegraphics[scale=0.32,angle=270]{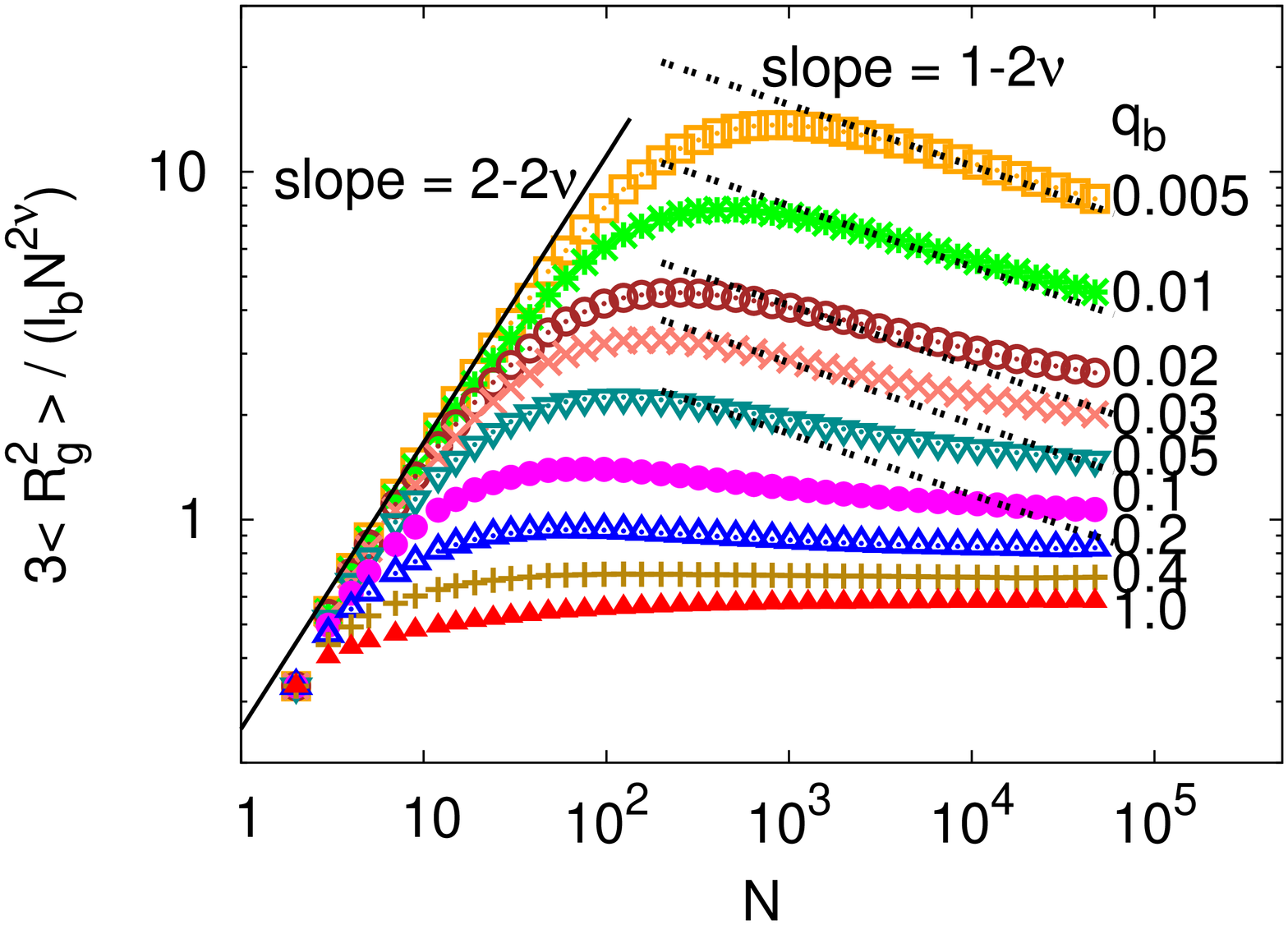}\hspace{0.4cm}
(b)\includegraphics[scale=0.32,angle=270]{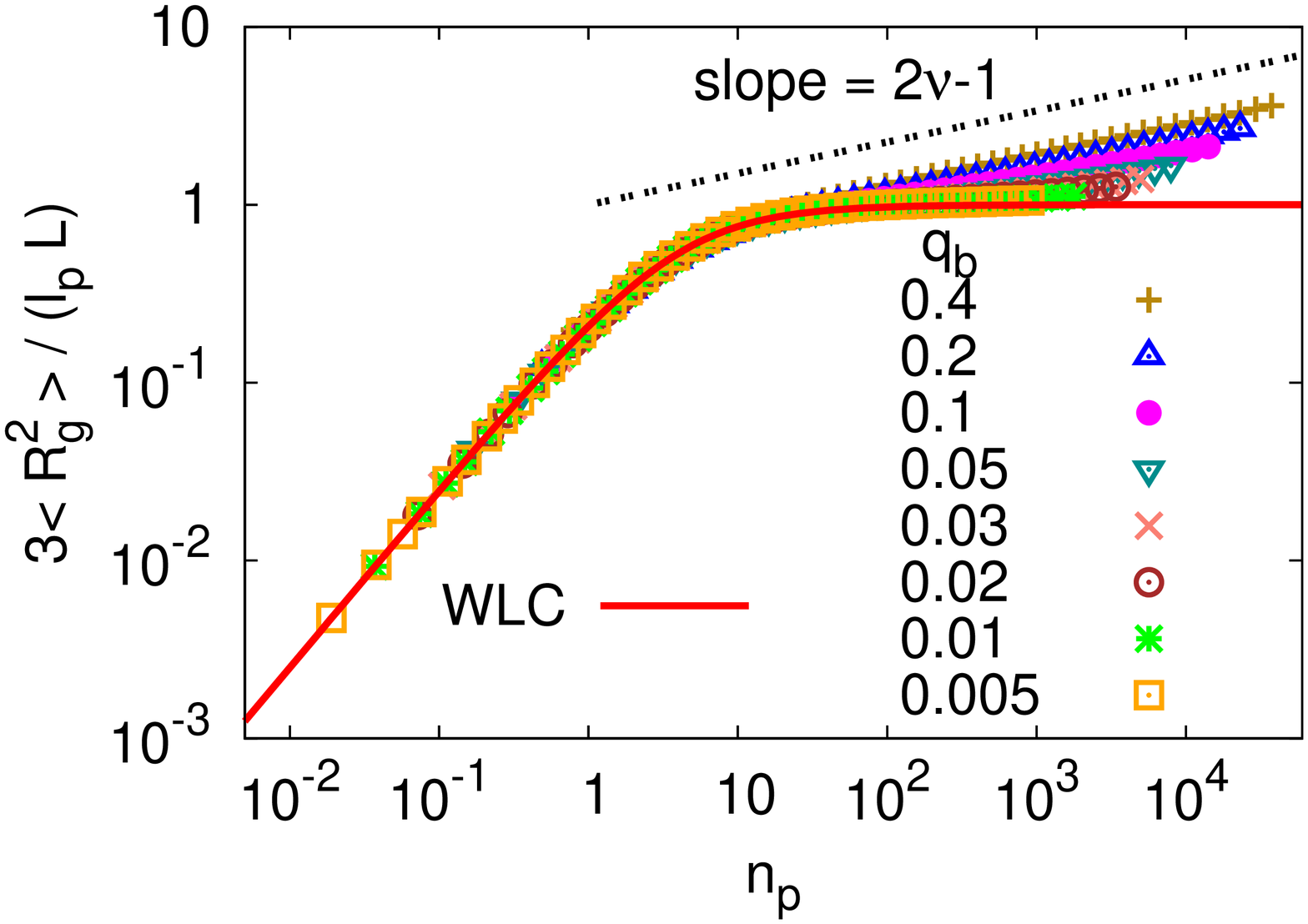}\\
(c)\includegraphics[scale=0.32,angle=270]{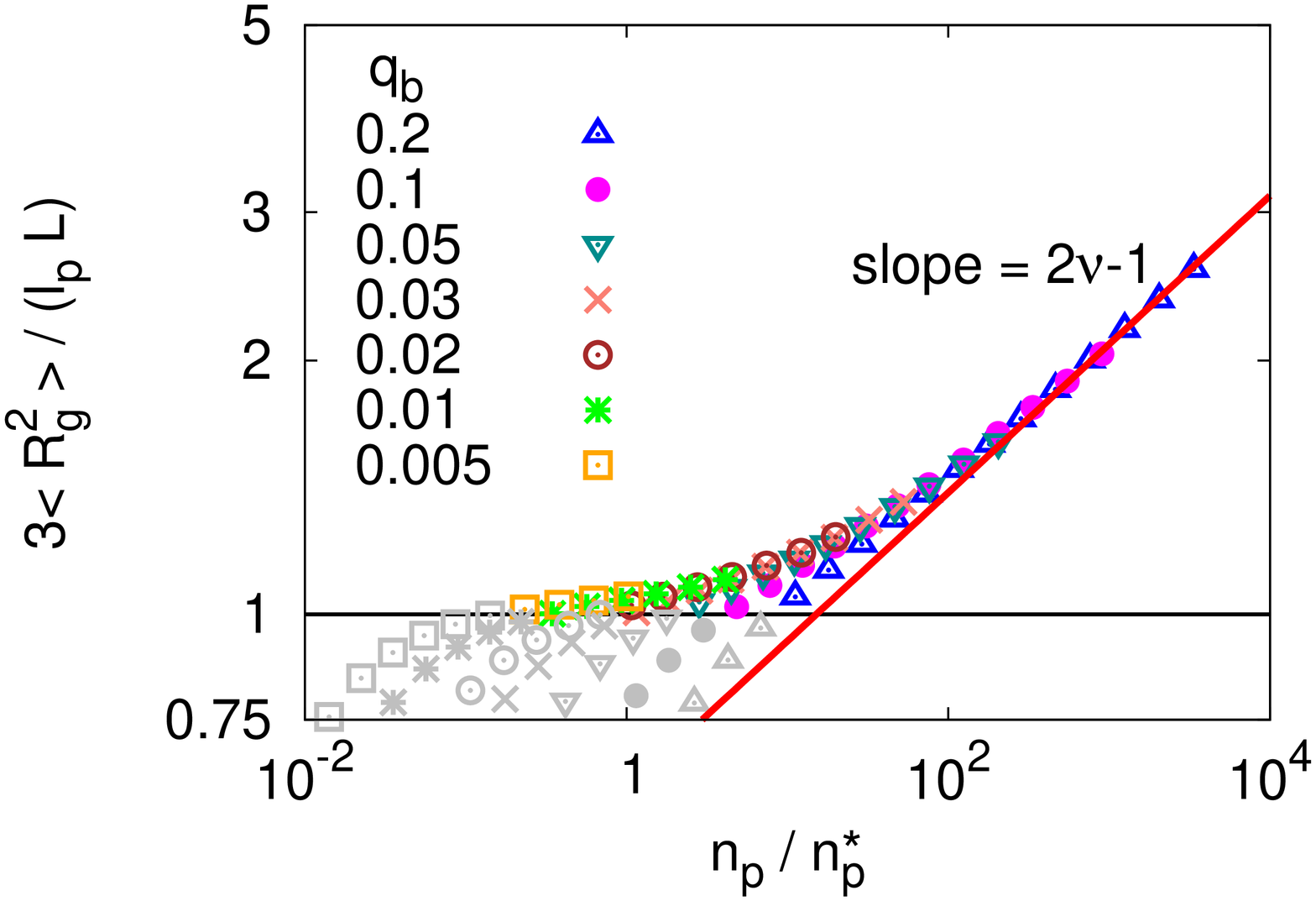}\hspace{0.4cm}
\caption{\label{fig7} (a) Log-Log plot of the relaxed mean square
gyration radius $3 \langle R^2_g \rangle / \ell_b N^{2 \nu}$ versus $N$,
for chain lengths $N$ up to $N=50 000$, and many values of the
stiffness parameter $q_b$, as indicated. The straight line with
slope $2-2\nu$ shows the slope reached for small $N$ in the rod-like
regime, where $\langle R^2_g \rangle= \ell^2_b N^2 /12$; the
straight dotted line with slope $1-2\nu$ (for intermediate values of $N$)
indicates the behavior expected for Gaussian chains,
$\langle R^2_g \rangle =\ell_b \ell_p N/3$.
(b) Log-log plot of $3 \langle R^2_g \rangle / \ell_p L$ versus
$n_p= N \ell_b/\ell_p$, using the same data as in (a) to test
the K-P model (full curve). The different choices of $q_b$ are shown
by different symbols, as indicated. The slope that one expects for all
$q_b$ for $N \rightarrow \infty$ ,  $(2 \nu -1)$, is indicated by a
broken straight line.}
\end{center}
\end{figure}

\section{SIMULATION RESULTS}
\subsection{Mean square gyration radii and their analysis}

We start with a description of our results for the mean square gyration 
radius $\langle R^2_g \rangle$ of the semiflexible SAW model (similar data 
for the mean square end-to-end distance $\langle R^2 \rangle$ of this model 
have already been presented elsewhere~\cite{68,69}), Fig.~\ref{fig7}. 
We clearly see that there are three regimes (Fig.~\ref{fig7}a): 
in the chosen normalization where we divide out the asymptotic power law 
$\langle R^2_g \rangle \propto N^{2 \nu}$, we first have a regime where 
$3 \langle R^2_g \rangle / \ell_b N^{2 \nu}$ increases with $N^{2-2\nu}$. 
For small $q_b$ this is interpreted as a rod-like regime; for $q_b \geq 0.2$ 
the chains are still too flexible, however, so a strictly rod-like behavior 
cannot yet be seen. Then a maximum occurs, and the ratio 
$3 \langle R^2_ g \rangle/ \ell_b N^{2\nu}$ decreases, before it settles 
down, after a second smooth crossover at a horizontal plateau (which 
according to Eq.~(\ref{eq14}) defines the value $\ell_p^{R_g}$). 
While this plateau for $q_b \geq 0.05$ is (presumably) actually reached 
for $N=50000$, the data also indicate that for $q_b \leq 0.03$ even chains 
of length $N$=50000 are at least an order of magnitude too short to allow 
a direct convincing estimation of the amplitude value $\ell_p^{R_g}$. 
On the other hand, even for $q_b=0.005$ (where we estimate from 
Eq.~(\ref{eq4}) that the persistence length $\ell_p$ is as large as 
$\ell_p \approx 52$~\cite{60}) the slope of the data in the intermediate 
regime has not fully reached the theoretical value $1-2 \nu$, the slope of 
the data in Fig.~\ref{fig7}a is still affected by crossover effects: 
the gradual crossover away from the Gaussian plateau towards the excluded 
volume-dominated behavior already starts when the gradual crossover from 
the rod-like regime to the Gaussian regime ends. 
Thus, even stiffer chains would 
be required to have a fully developed Gaussian behavior of the gyration 
radius. Fig.~\ref{fig7}b now attempts a scaling plot, where the 
persistence length estimates extracted from Eq.~(\ref{eq4}) were used 
to rescale $\langle R^2_g \rangle$ in the K-P model representation 
(cf. Eq.~(\ref{eq17})). It is evident that the rod-like regime and 
the onset of the crossover towards the K-P plateau are very well described 
by Eq.~(\ref{eq17}). For $q_b \geq 0.2$, of course, there is basically 
a direct crossover from the rod-like regime to the excluded volume 
dominated regime, but even then it is evident that the curves do not 
superimpose on a master curve, as they do in $d=2$ dimensions~\cite{60,64}, 
but rather splay out systematically, and the smaller $q_b$ becomes 
(and hence the larger $\ell_p$ becomes) the more the data still are 
slightly above the K-P plateau.

Using the estimates for $n^*_p (q_b)$ extracted from the analysis of 
$\langle R^2 \rangle$ for this model in our previous work~\cite{68,69}, 
the data for $n_p \gg n^*_p$ do collapse on a simple straight line on the 
log-log plot, however (Fig.~\ref{fig7}c). For $n_p$ near $n^*_p$ the 
curves splay out, the master curve describing this second crossover from 
the K-P plateau to the excluded volume power law emerges as an envelope of 
the curves for individual values of $q_b$ (which fall increasingly
below the master curve 
in the crossover region the larger $q_b$ is). Of course, 
$n^*_p \propto (\ell_p/\ell_b)^\zeta$ cannot produce a scaling of 
the crossover towards the rod-behavior, there the curves must splay out, 
irrespective of how small $q_b$ is, but the deviation of the data from 
the horizontal K-P plateau moves more and more to the left of the plot the 
smaller $q_b$ becomes.

\begin{figure}
\begin{center}
% for a multi-line caption
%\onelinecaptionstrue
% for a one-line caption
%\includegraphics{}
\includegraphics[scale=0.32,angle=270]{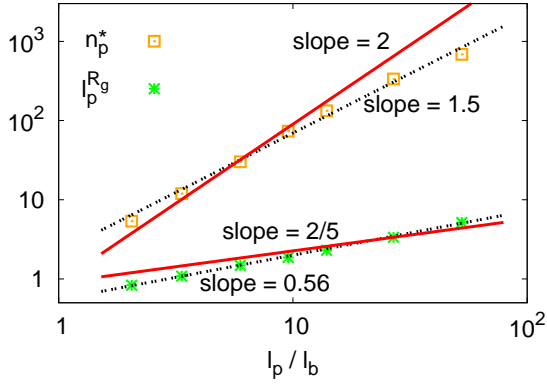}\hspace{0.4cm}
\caption{\label{fig8} Log-log plot of $n^*_p$ and $\ell^{R_g}_p /\ell_b$,
as indicated in the figure, versus $\ell_p/\ell_b$, using data for
$q_b=0.2.$ to $0.005$ (left to right). Dotted lines indicate the observed exponents
while full straight lines show the Flory predictions for the exponents.}
\end{center}
\end{figure}

\begin{figure}
\begin{center}
% for a multi-line caption
%\onelinecaptionstrue
% for a one-line caption
%\includegraphics{}
(a)\includegraphics[scale=0.32,angle=270]{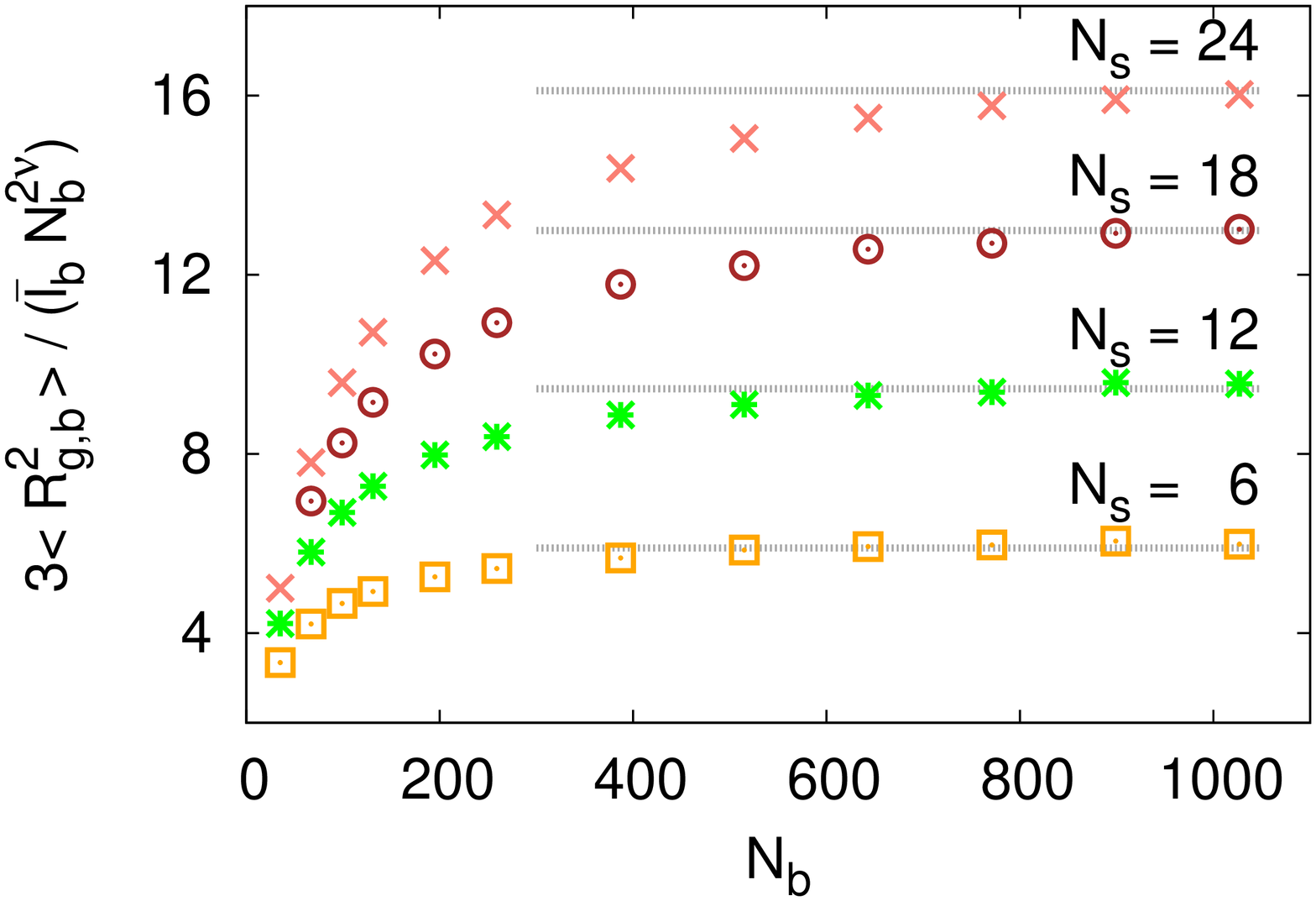}\hspace{0.4cm}
(b)\includegraphics[scale=0.32,angle=270]{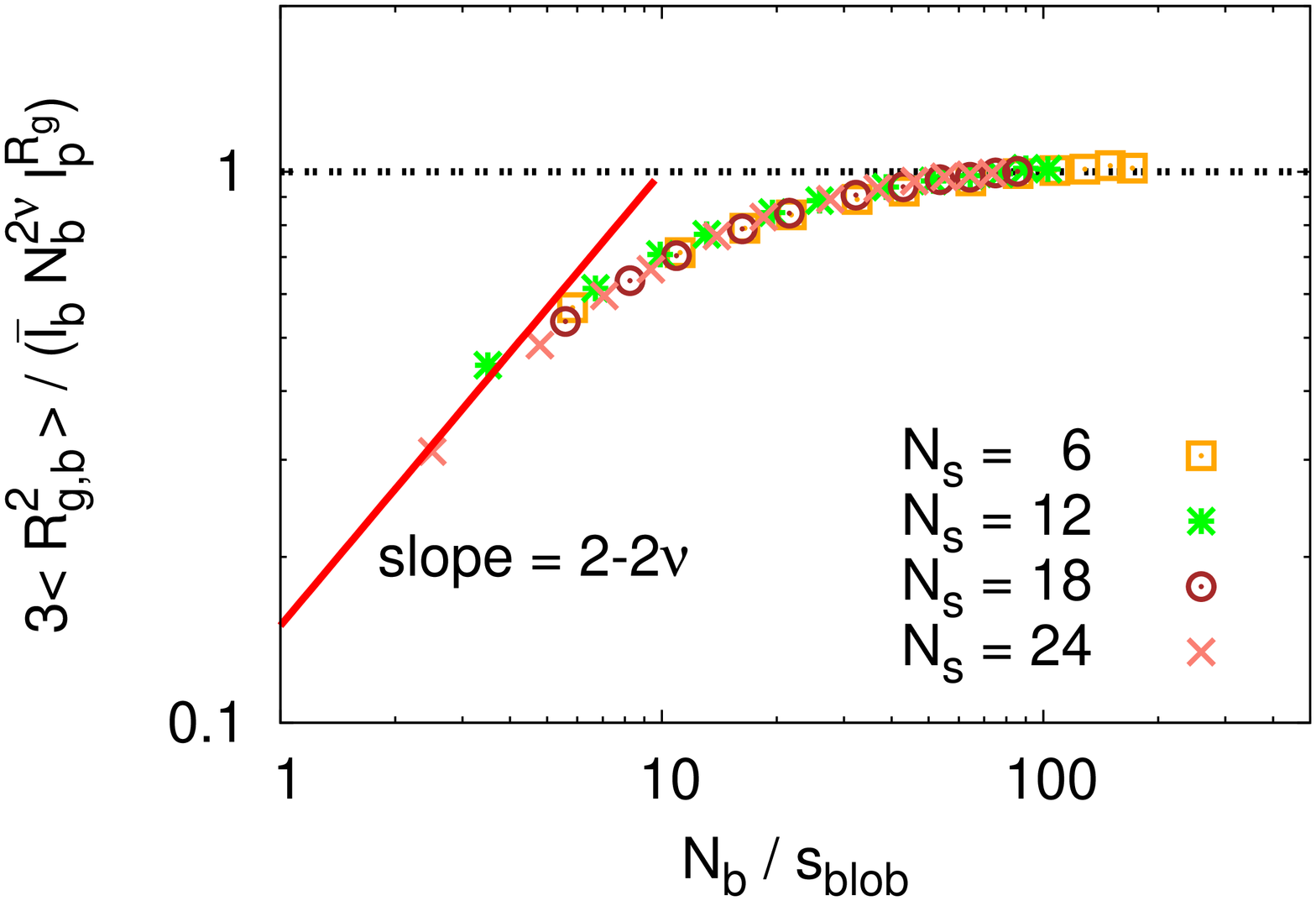}\\
\caption{\label{fig9} (a) Plot of the normalized mean square
gyration radius of the backbone of the model for bottle-brush
polymers, $3 \langle R^2_{g,b} \rangle / \bar{\ell}_b N^{2 \nu}$,
versus the backbone chain length $N_b$, for four different side chain
lengths $N_s$, $N_s=6$, $12$, $18$, and $24$, as indicated. Here
$\bar{\ell}_b \approx 2.7$ is the average bond length of the single
bond in the bond fluctuation model under good solvent conditions.
The horizontal plateaus allow to extract the estimates for
$\ell^{R_g}_p$ defined in Eq.~(\ref{eq14}). (b) Same data as in (a),
but ordinate is rescaled with $\ell^{R_g}_p$ so for
$N_b \rightarrow \infty $ all data converge to one, and abscissa
is rescaled by the effective blob-size $s_{\rm blob}$ (see text),
on a log-log plot. Straight line indicates the rigid-rod behavior,
with slope $2-2\nu$ in this representation.}
\end{center}
\end{figure}

\begin{figure}
\begin{center}
% for a multi-line caption
%\onelinecaptionstrue
% for a one-line caption
%\includegraphics{}
(a)\includegraphics[scale=0.32,angle=270]{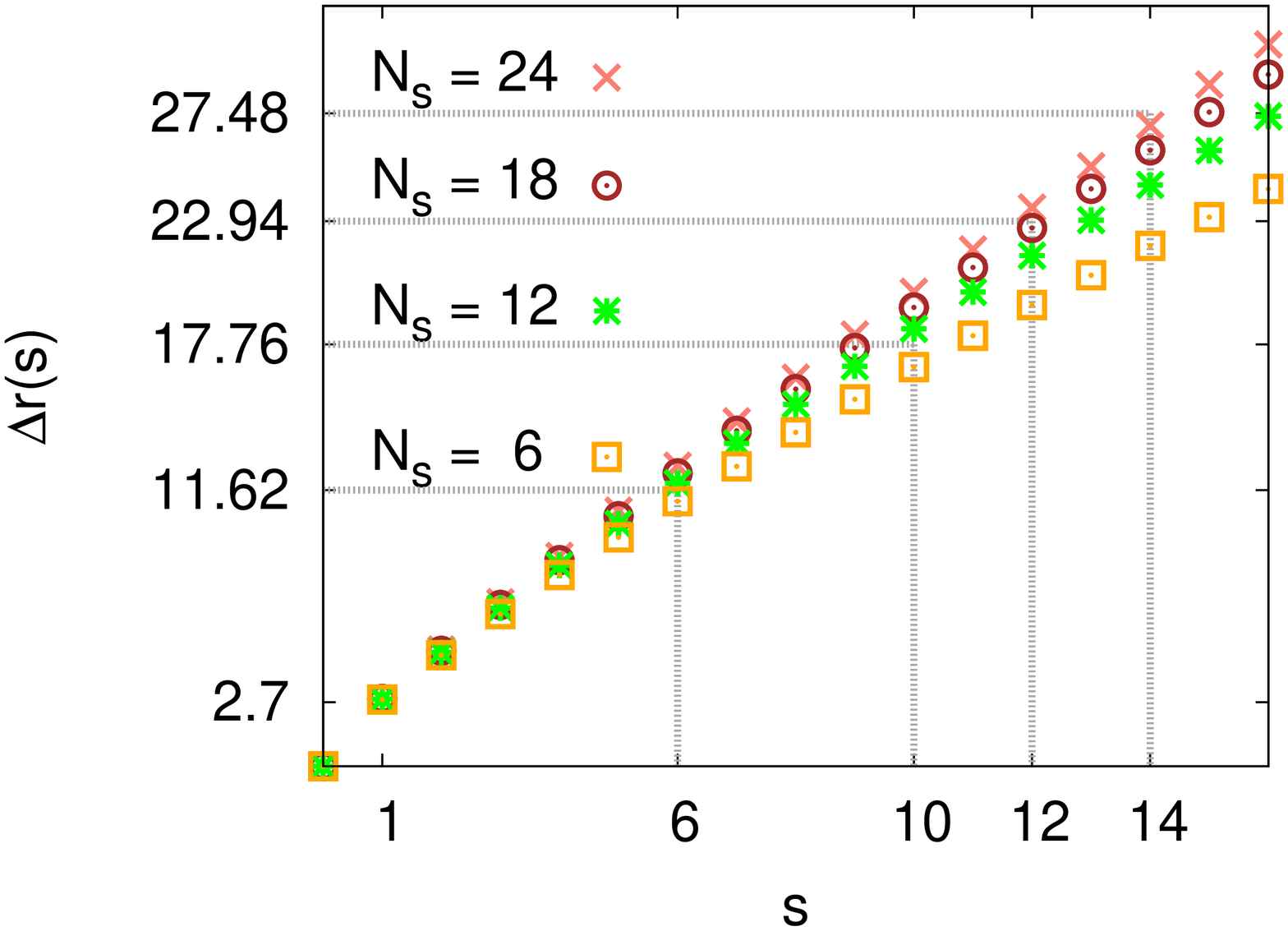}\hspace{0.4cm}
(b)\includegraphics[scale=0.32,angle=270]{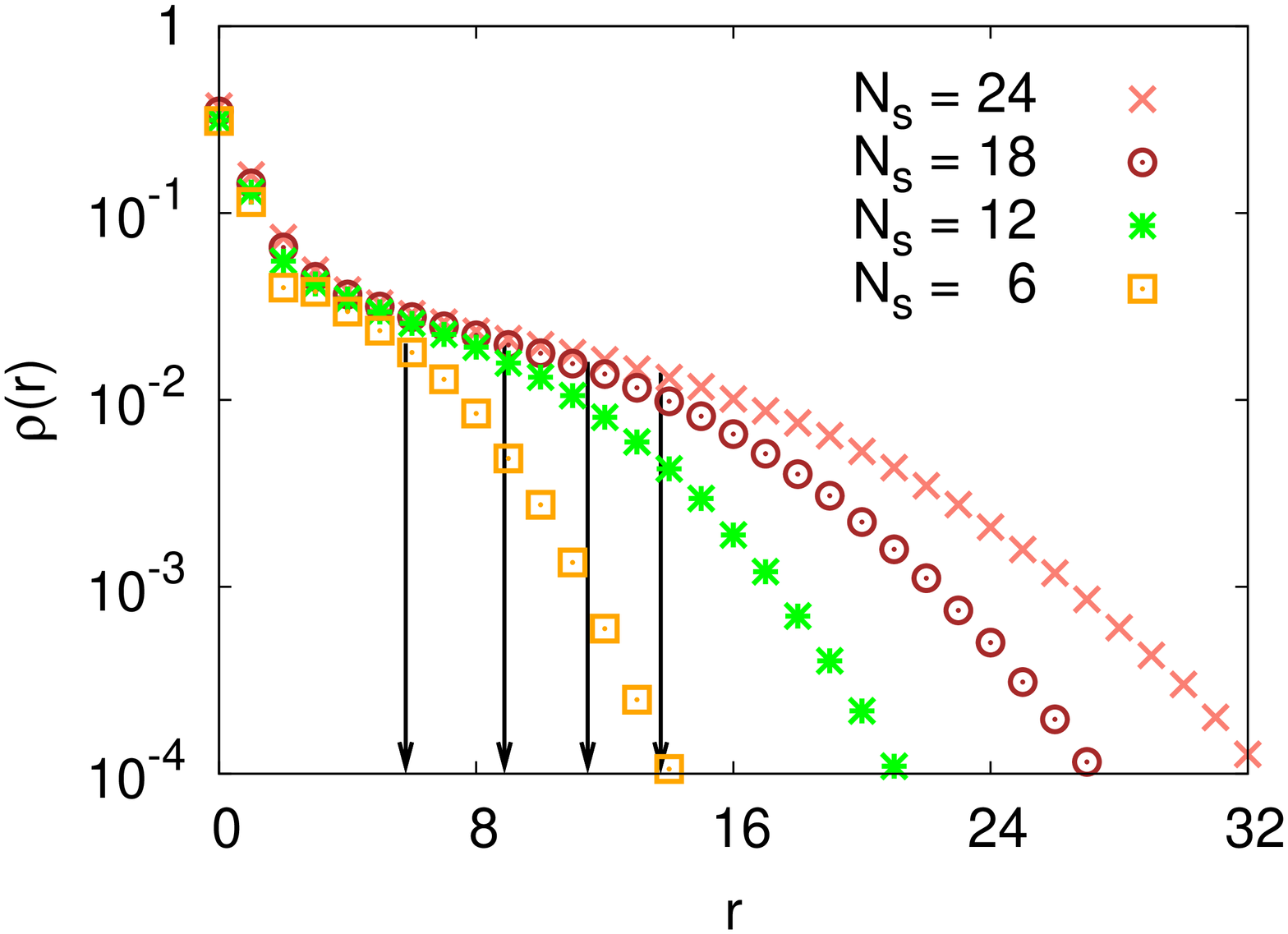}\\
\caption{\label{fig10} (a) End-to- End distance $\Delta r(s)$ of
subchains containing $s$ successive backbone monomers for
$N_s=6$, $12$, $18$, and $24$. The horizontal solid lines and
the numbers shown on the ordinate indicate the
choices $\Delta r(s)= 2 R_{cs} (N_s)$, from which the corresponding
values of $s_{\rm blob}$ can be read off (vertical straight lines),
namely, $s_{\rm blob}=6$, $10$, $12$,
and $14$ for $N_s=6$, $12$, $18$ and $24$, respectively.
$R_{cs}(N_s)$ is the cross-sectional radius, which is extracted
from the radial monomer density profiles $\rho(r)$ as shown in (b).
(b) Radial monomer density profiles $\rho(r)$ in planes locally
perpendicular to the backbone of bottle-brush polymers with backbone
length $N_b=1027$ and plotted versus radial distance $r$ for side
chain lengths $N_s=6$, $12$, $18$ and $24$, as indicated.
The cross-sectional radius then follows as
$\langle R^2_{cs} \rangle = 2 \pi \int\limits^\infty_0 rdr \rho(r) r^2$
with the density profile being normalized as
$2 \pi \int\limits^\infty_0 rdr \rho(r)=N_s$. The values of $R_{cs}$ are
pointed out by arrows.}
\end{center}
\end{figure}

Recalling that for the semiflexible SAW model the effective chain 
thickness $D$ simply is $D=\ell_b=1$, the Flory theory, Eq.~(\ref{eq20}), 
simply predicts $n^*_p \propto \ell^2_p$; $\zeta=2$, while the rod to 
Gaussian coil behavior occurs around $n_p=1$, of course. Qualitatively, 
our data are in good agreement with these predictions, but not quantitatively: 
This is illustrated in Fig.~\ref{fig8}, where $n^*_p$ and $\ell^{R_g}_p$ 
are plotted in log-log form versus $\ell_p/\ell_b$. It is seen that instead 
of the theoretical value $\zeta=2$ an exponent $\zeta=1.5$ is observed. 
Now it is clear that Flory arguments imply also $\nu=3/5$ instead of 
$\nu \approx0.588$~\cite{7}, but this small difference cannot account 
for the large discrepancy encountered here. It would be desirable to study 
much larger values of $N$ to confirm whether this discrepancy is a real
effect (or our estimation of the crossover master curves in Fig.~\ref{fig7}c
for $\langle R_g^2 \rangle$ (and for $\langle R ^2 \rangle$ in~\cite{69}) 
are systematically off). 
Thus, more work is still needed to fully clarify the situation.

It turns out that the behavior of our model for the bottle-brush 
polymers (which can describe actual scattering data for bottle-brush 
polymers very well, as demonstrated by Hsu et al.~\cite{17}) is 
much simpler: a plot of the mean square gyration radius 
$\langle R^2_{g,b} \rangle$ of the backbone versus backbone chain length 
$N_b$, for different side chain lengths $N_s$, Fig.~\ref{fig9}a, 
normalized by $N^{2\nu}_b$ reveals a monotonic increase towards a plateau, 
there is not the slightest indication of a regime where the ratio 
$\langle R^2_{g,b} \rangle /N^{2\nu}_b$ decreases, unlike the behavior 
of the semiflexible SAW (Fig.~\ref{fig7}a).
Thus, there is no evidence
whatsoever for a Gaussian K-P plateau for this model. But the increase of 
the plateau value $\ell^{R_g}_p$ with increasing side chain length $N_s$ 
does indicate that the chain considerably stiffens, as $N_s$ increases. 
However, this stiffening goes along with an increase in the effective 
chain thickness $D$. The latter can be estimated from the radial density 
profile (Fig.~\ref{fig10}) by identifying the diameter $D$ of the 
bottle-brush as $D=2 R_{cs} (N_s)\equiv 2 \sqrt{\langle R^2_{cs} \rangle }$. 
Hsu et al.~\cite{68,69} suggested to coarse-grain the bottle-brush along 
the backbone, dividing it into ``blobs'' along the chemical sequence of 
the backbone. The chemical distance $s_{\rm blob}$ along the backbone 
between its exit and entry points into a blob is found from a simple 
construction which assumes that the blobs are essentially spherical, 
so the geometrical distance $\Delta r(s)$ between exit and entry points 
of the backbone should be equal to $D$. Recording $\Delta r(s)$ for 
arbitrary $s$, Fig.~\ref{fig10}a, using the equation 
$\Delta r (s_{\rm blob})=D$ allows us to simply read off the numbers 
$s_{\rm blob}$ for the choices of $N_s$, as illustrated in 
Fig.~\ref{fig10}. The success of the rescaling shown in 
Fig.~\ref{fig9}b shows that the persistence length $\ell_p$ of 
bottle-brushes simply is proportional to $D$. We also recognize that 
the asymptotic SAW-like behavior (where the horizontal plateau 
in Fig.~\ref{fig9} is reached) only occurs for about 
$N_b/s_{\rm blob} \approx 60.$ Comparing this behavior to 
Fig.~\ref{fig7}b, we see that there the power law (for $q_b=0.4$) or 
K-P plateau (for small $q_b$) is reached for 
$n_p= N \ell_b/\ell_p \approx 40$. Roughly, these successive blobs then are 
equivalent to one persistence length. This comparison suggests that for 
the bottle-brushes we should identify $\frac{2}{3} (N_b/s_{\rm blob})$ 
with $n_p$, i.e. the number of monomers along the backbone 
corresponding to one persistence length is  
$\frac{3}{2} s_{\rm blob}=9$, $15$, $18$ and $21$ for $N_s=6$, $12$, $18$ and 
$24$, respectively. Noting that the average bond length $\bar{\ell}_b$ 
in the bond fluctuation model is $\bar{\ell}_b=2.7$, we would obtain 
persistence lengths $\ell_p(N_s) = \frac{3}{2} D \approx 17$, $27$, $34$ and $41$ 
for $N_s=6$, $12$, $18$ and $24$, respectively. The result that 
$\ell_p(N_s)$ is of the same order as $D$ irrespective of the side 
chain length agrees with early theoretical predictions~\cite{18,19} 
but is at variance with the result of 
Fredrickson~\cite{20} who predicted a much faster increase of $\ell_p$ 
with $N_s$. However, Feuz et al.~\cite{30} pointed out that the result 
of Fredrickson~\cite{20} can only be expected to hold for extremely 
long side chains, such as $N_s=1000$. Such long side chains are 
neither relevant for simulations nor for experiment, however. 
We stress that the range of $N_s$ accessible to simulations 
(Figs.~\ref{fig9},~\ref{fig10}) nicely corresponds to the range of 
studied experimentally~\cite{28,32,33,34,35,36,37,38}.

The mapping performed in Fig.~\ref{fig9} means that we have 
coarse-grained the bottle-brush polymers (Fig.~\ref{fig4}a) 
into an effective bead-spring model (Fig.~\ref{fig4}d). If this mapping 
is taken literally, it can also be used to obtain the resulting 
coarse-grained contour length $L_{cc}$ (Fig.~\ref{fig3}) as
\begin{equation} \label{eq37}
L_{cc}= 2 R_{cs} (N_s) N_b/s_{\rm blob} (N_s) \,.
\end{equation}
Instead of the ``chemical'' contour length $L=N_b\ell_b \approx 2773$ 
a reduced length is found, namely $L_{cc} \approx 1989$,
$1824$, $1963$ and $2016$, for $N_s=6$, $12$, $18$ and $24$, respectively. 
This means that the coarse-grained contour length $L_{cc}$ is 
about $30\%$ smaller than the ``chemical'' contour length in this model.

\begin{figure}
\begin{center}
% for a multi-line caption
%\onelinecaptionstrue
% for a one-line caption
%\includegraphics{}
(a)\includegraphics[scale=0.32,angle=270]{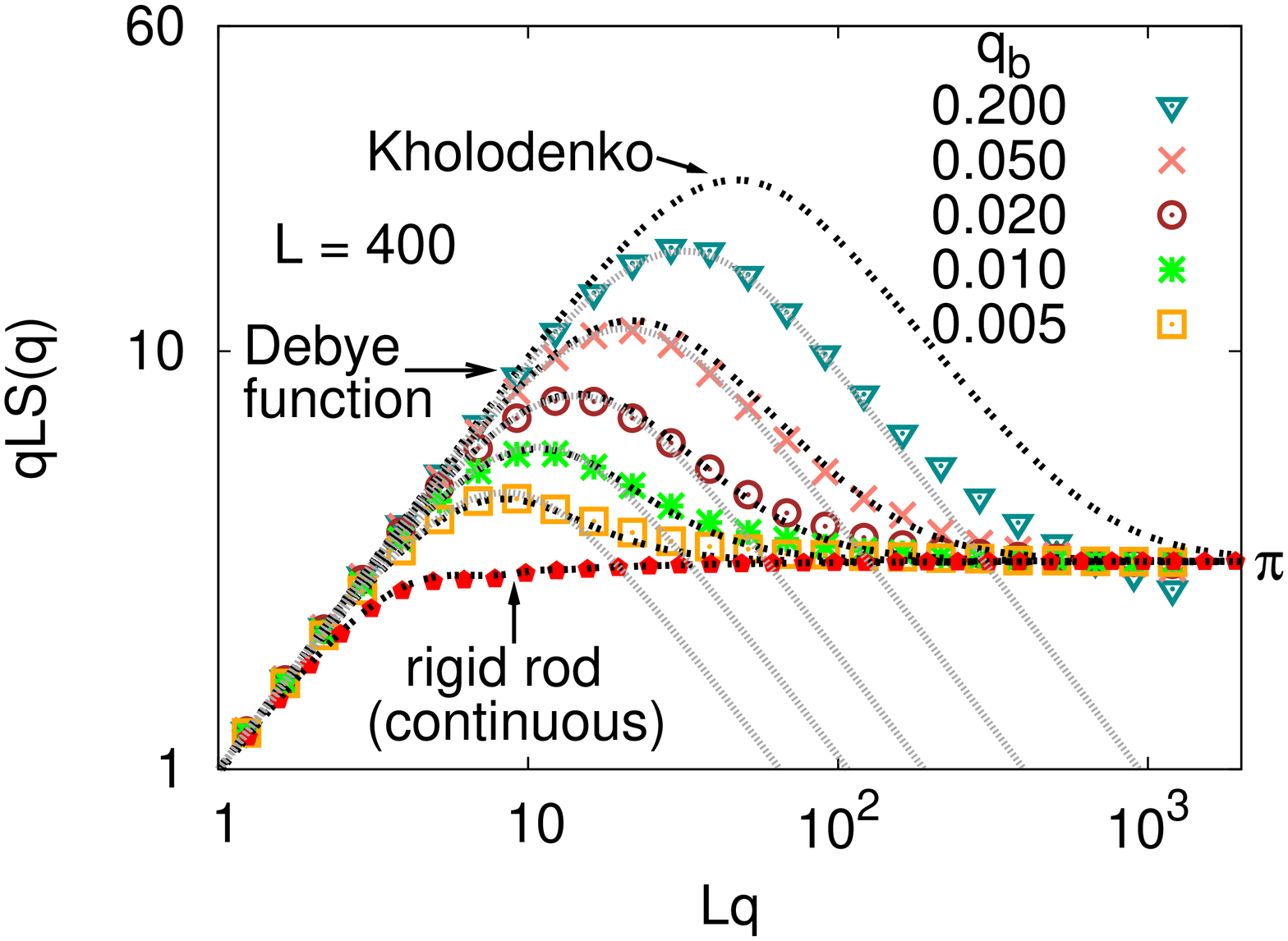}\hspace{0.4cm}
(b)\includegraphics[scale=0.32,angle=270]{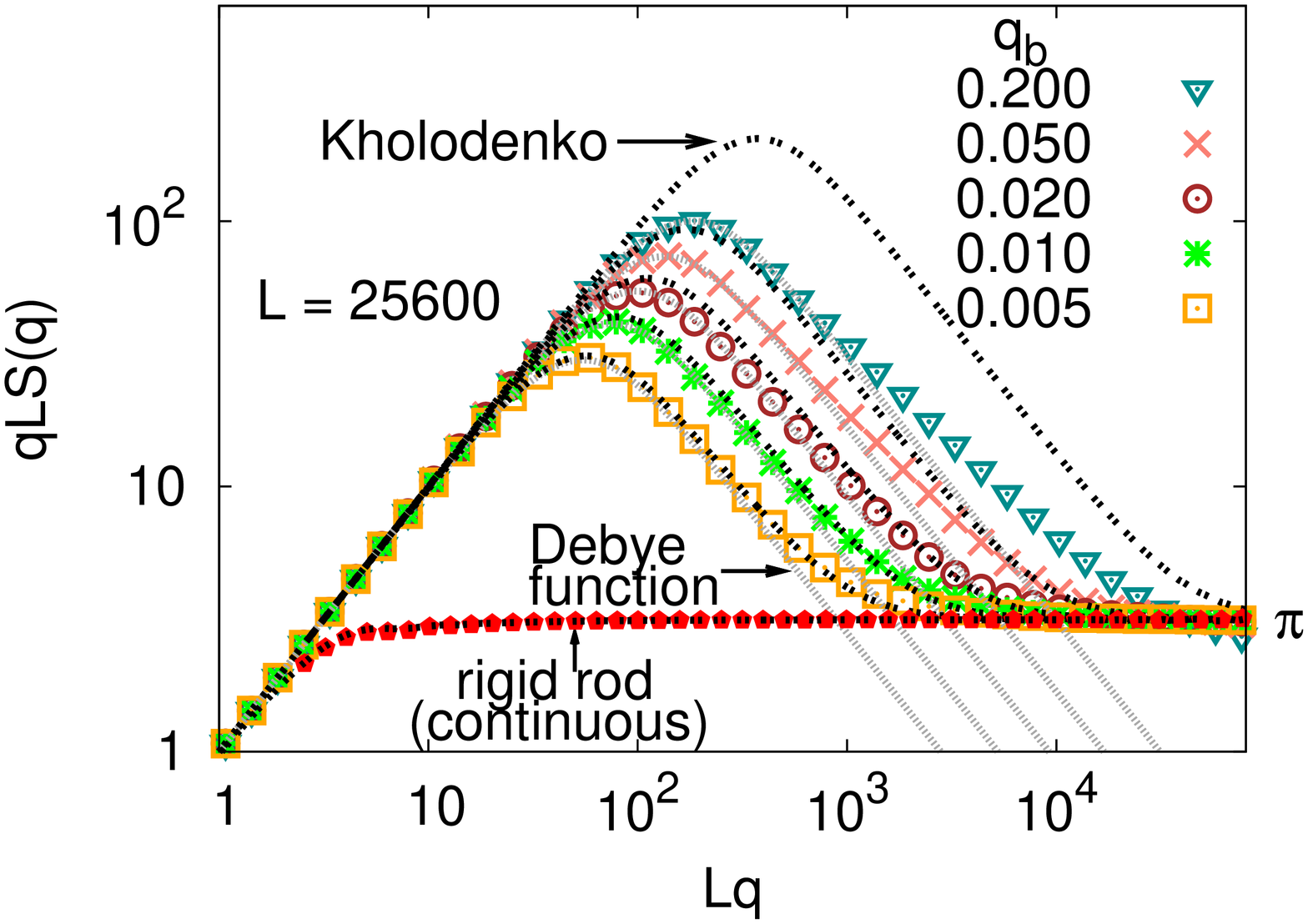}\\
\vspace{0.4cm}
(c)\includegraphics[scale=0.32,angle=270]{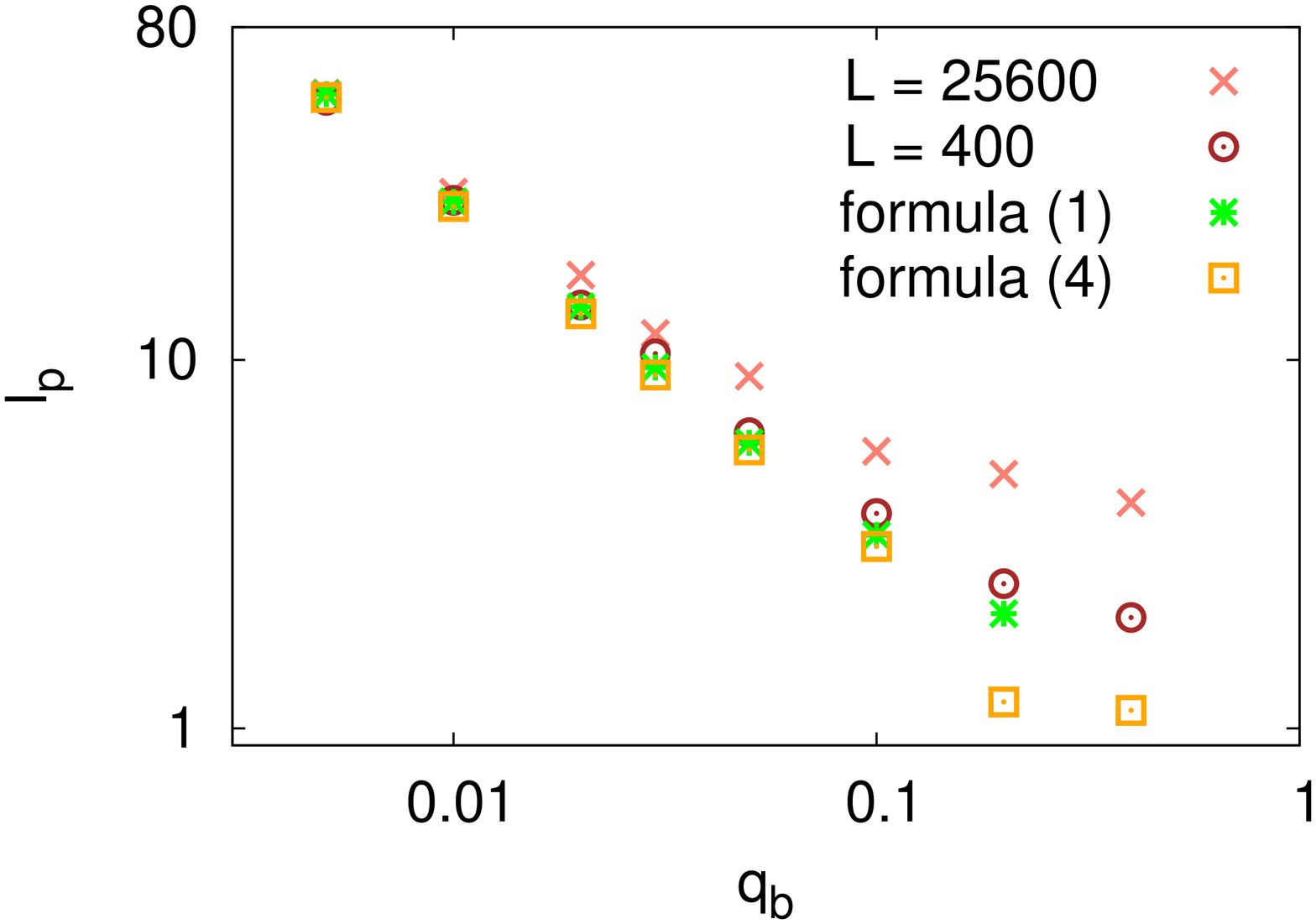}
\caption{\label{fig11} Kratky plot of the structure factor of the
semiflexible self-avoiding walk model, log-log plot of $q LS(q)$
versus $q L$, for two chain lengths, namely $L=400$ (a) and $L=25600$ (b).
(c) Persistence length $\ell_p$ plotted vs.~$q_b$ for $0.005\le q_b \le 0.4$.
In (a)(b) several choices of $q_b$ are included, namely
$q_b=0.2$, $0.05$, $0.02$, $0.01$, $0.005$, as indicated.
The scattering functions of a rigid rod and the Debye function
are included, as well as the prediction of
Kholodenko \{Eqs.~(\ref{eq28})-(\ref{eq30})\}.
The predicted large $q$-limit of $\pi$ is indicated.
In the Debye function the observed value of $\langle R^2_g \rangle$
was used as an input, while for the Kholodenko formula the
persistence length (estimated from Eq.~(\ref{eq4})) was used
as an input. In (c) the persistence length is taken as the best fitting
parameter such that the prediction of Kholodenko formulas describes the
correct maximum in the Kratky plot for our simulation data of chain lengths
$L=400$, and $25600$.
The estimates using Eqs.~(\ref{eq1}), (\ref{eq4}) are also shown
in (c) for comparison.}
\end{center}
\end{figure}

\begin{figure}
\begin{center}
% for a multi-line caption
%\onelinecaptionstrue
% for a one-line caption
%\includegraphics{}
\includegraphics[scale=0.32,angle=270]{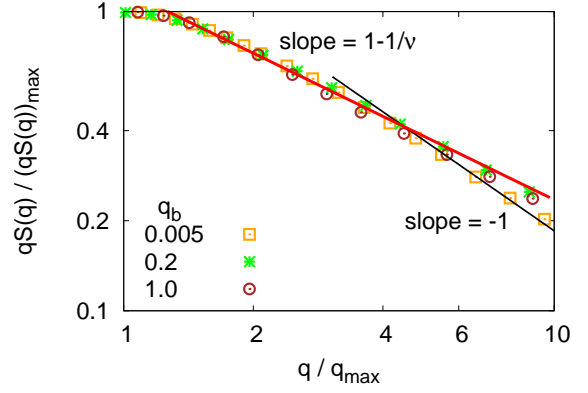}\hspace{0.4cm}
\caption{\label{fig12} Kratky plot for $q_b=1.0$, $0.2$, and $0.005$
plotted in a rescaled form, $q S(q)/(q S(q))_{\rm max}$ vs.~$q/q_{\rm max}$,
restricting the ordinate range to the decade from
1 to 0.1 and the abscissa to the range from 1 to 10. The theoretical
power laws (slope$=1-1/\nu \approx 2/3$ and slope $=-1$,
respectively, as predicted in Fig.~5) are included.}
\end{center}
\end{figure}

\subsection{Analysis of the structure factor}

We now turn to the structure factor of the semiflexible SAW
model presenting Kratky plots for 
two contour lengths, $L=400$ and $L=25600$, in Fig.~\ref{fig11}. 
As expected, cf. Fig.~\ref{fig5}, one first has a linear increase 
with $qL$, then a round maximum followed by a decrease which then 
gradually settles down at a horizontal plateau, that again is 
compatible with the theoretical prediction, $\pi$.

While for the short chain length ($L=400$) the agreement with the 
theoretical prediction (due to Kholodenko~\cite{87}, 
Eqs.~(\ref{eq28})-(\ref{eq30}), which were found~\cite{95} to be 
numerically almost indistinguishable from the exact result provided by 
Stepanow~\cite{93,94}) is almost perfect, for the very long chains 
$(L=25600)$ we note systematic deviations between Kholodenko's 
prediction~\cite{87} and the data for relatively large $q_b$ near 
the maximum of the Kratky plot. This must be expected, since the 
input in the Kholodenko formula is just the persistence length 
$\ell_p$ \{which we have extracted from Eq.~(\ref{eq4})\} 
and implicit in the theory is the Gaussian prediction for 
$\langle R^2_g \rangle$, namely $\langle R^2_g \rangle = L \ell_p/3$ 
\{Eq.~(\ref{eq13})\}. As seen in Fig.~\ref{fig7}, for $q_b=0.2$ 
already rather clear deviations from this result occur for $N=25600$, 
while for small $N$ such as $N=400$ such deviations still are rather 
small. In contrast, in the Debye formula the correct (as observed) 
value of $\langle R^2_g \rangle$ was used as an input: then deviations 
from the Debye function are only seen near the region where the 
crossover to $q LS(q) = \pi$ starts to set in at large $q$ 
(the Debye function does not describe this crossover at all). 
Since the shape of the Kholodenko function always is rather similar 
to the actual function, it is obvious that one always can fit the 
data to the Kholodenko function, if $\ell_p$ is not known: 
however, the resulting fitted persistence length will be 
systematically too large, if excluded volume effects are present 
as shown in Fig.~\ref{fig11}c.

To elucidate the significance of excluded volume on the structure factor 
further, we show a magnification of the region near the maximum for 
$q_b=1$, $0.2$, and $0.005$ in Fig.~\ref{fig12}. It is seen that the 
identification of the two power laws suggested for the decay of $q S(q)$ 
in the region beyond the maximum of the Kratky plot is rather subtle. 
In particular, for rather stiff chains the crossover to the rod-like 
scattering sets in rather early, so for the clear identification of 
power laws the available range of $q$ simply is not large enough. 
This very gradual crossover between the three different regimes 
(rods to Gaussian coils to coils swollen by the excluded volume 
interaction) complicates the data analysis, if only a restricted 
range of parameters (such as the chain length $N$ and the wavenumber 
$q$) can be investigated.

\begin{figure}
\begin{center}
% for a multi-line caption
%\onelinecaptionstrue
% for a one-line caption
%\includegraphics{}
\includegraphics[scale=0.32,angle=270]{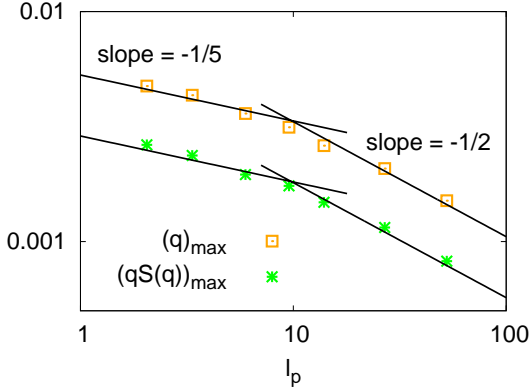}\hspace{0.4cm}
\caption{\label{fig13} Log-log plot of $q_{\rm max}$ and
$(q S(q))_{\rm max}$ versus $\ell_p$ (as estimated
using Eq.~(\ref{eq4})) using data for $q_b=0.2$ to
$0.005$. Straight lines indicated the exponents
$q_{\rm max} \propto \ell^{-1/5}_p$ and $q_{\rm max} \propto \ell^{-1/2}_p$
that one expects according to the Flory treatment in the excluded
volume region and Gaussian region, respectively.
All data were taken for N=50000.}
\end{center}
\end{figure}

The smoothness of the crossover also becomes evident when one 
studies the dependence of the  position of the peak in the Kratky-plot 
(and its height) on the persistence length (Fig.~\ref{fig13}). 
Typically, the data fall neither in the regime where strict 
Gaussian behavior occurs, nor in the regime where excluded 
volume scaling is fully developed.

Despite all these difficulties due to the gradual crossovers, 
the semiflexible SAW nevertheless
is a relatively simple case, since one knows 
that here $D=\ell_b(=1)$, and $\ell_p$ can be varied over a wide range 
by variation of $q_b$, keeping all other parameters constant, and 
moreover $\ell_p$ can be estimated precisely from the initial decay 
of the bond vector autocorrelation function (or, equivalently, from 
Eq.~(\ref{eq4})). For the second model studied here, bottle-brush 
polymers under good solvent conditions, we have seen that varying the 
side chain length $N_s$ we change $D$ and $\ell_p$ together, and also 
the coarse-grained contour length $L_{cc}$ is significantly smaller 
than the chemical contour length $N_b \bar{\ell}_b$, and it is 
nontrivial to estimate $L_{cc}$ accurately.

\begin{figure}
\begin{center}
% for a multi-line caption
%\onelinecaptionstrue
% for a one-line caption
%\includegraphics{}
(a)\includegraphics[scale=0.32,angle=270]{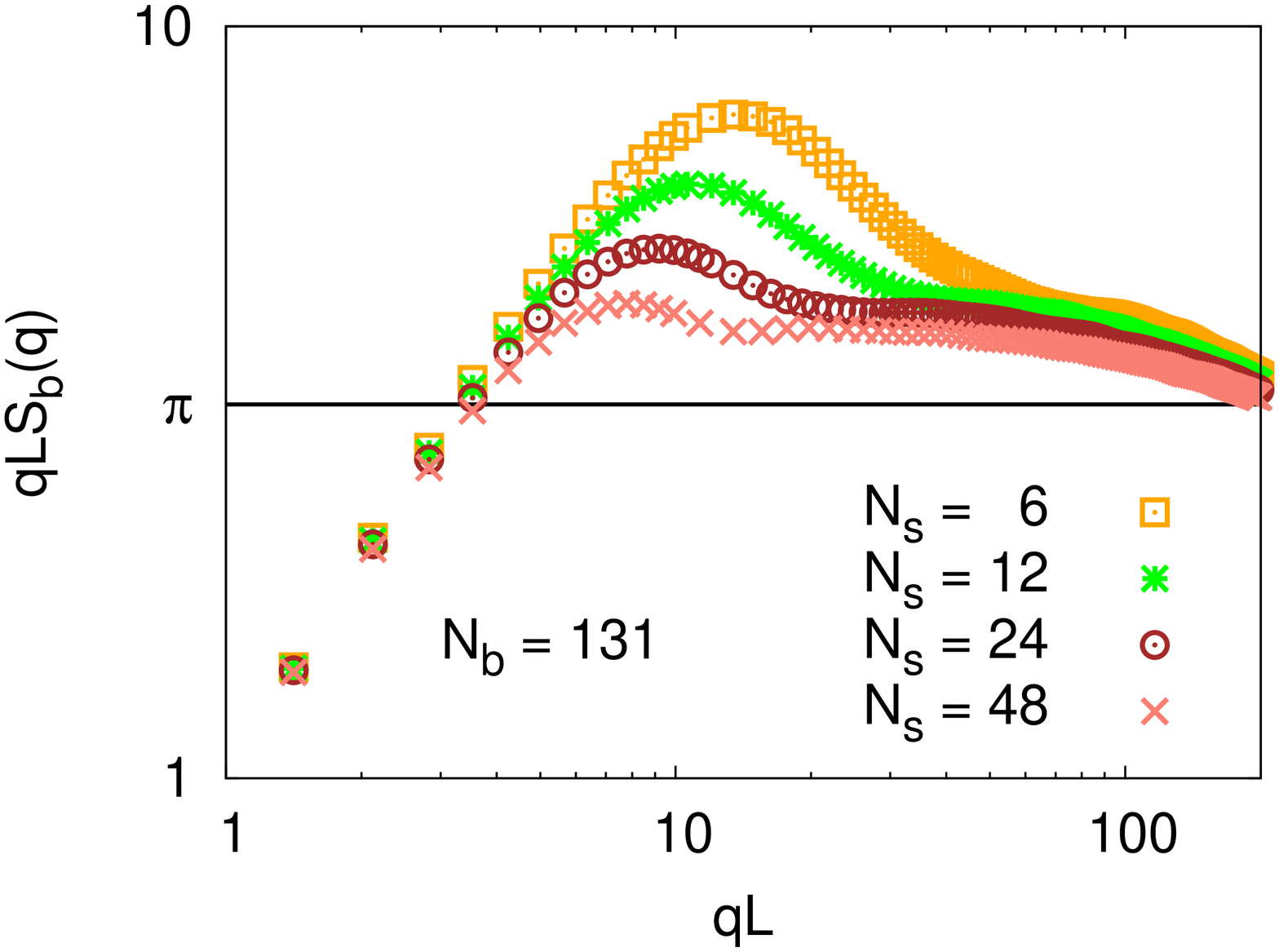}\hspace{0.4cm}
(b)\includegraphics[scale=0.32,angle=270]{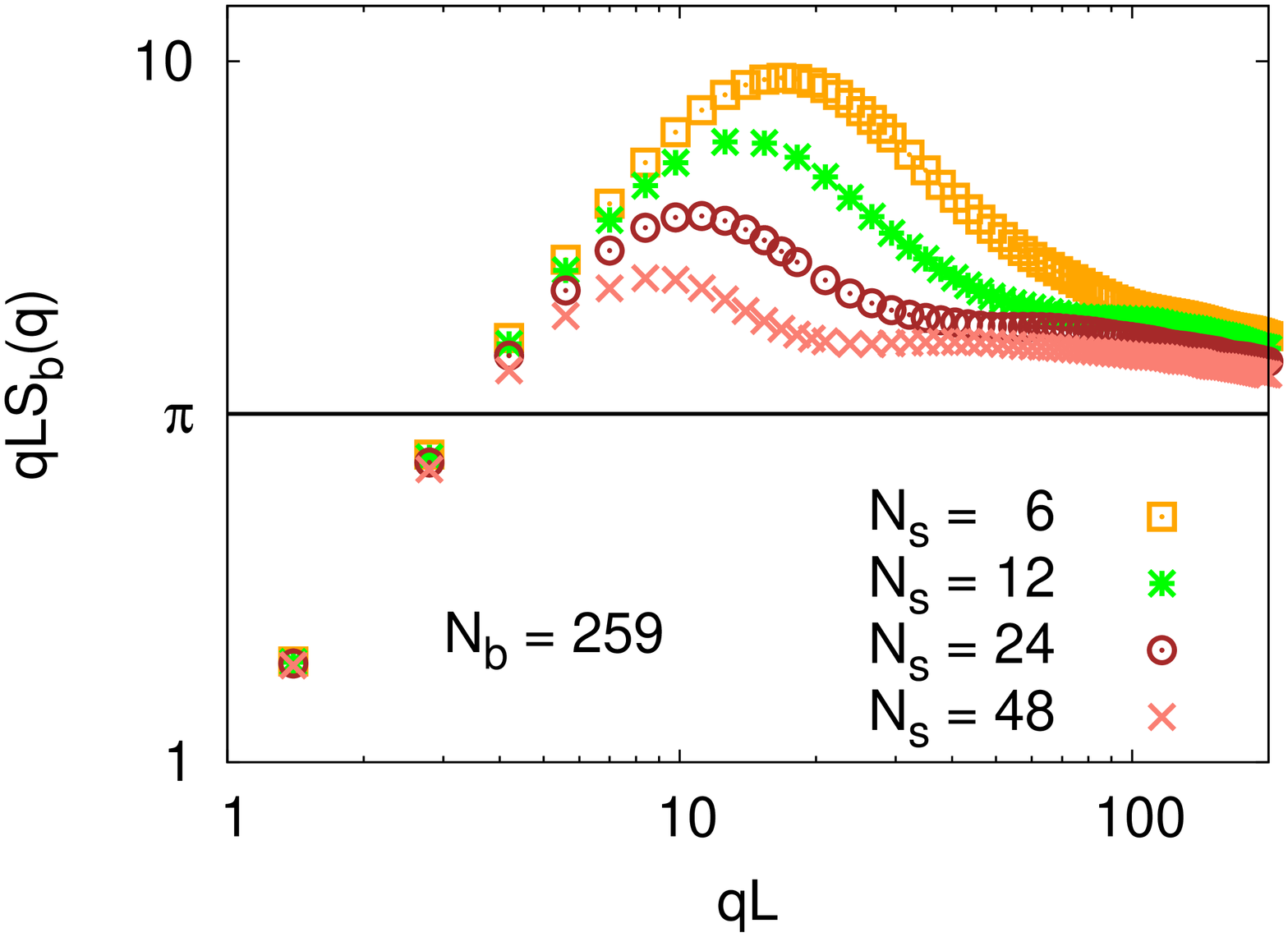}\\
\caption{\label{fig14} Kratky plot of $q L S_b(q)$ versus $qL$ for
bottle-brush polymers with $N_b=131$ (a) and $N_b=259$ (b),
where ``$L=N_b \bar{\ell}_b$'' is the ``chemical'' contour length and
$S_b(q)$ is the scattering function of the backbone only.
Four side chain lengths $N_s=6$, $12$, $24$ and $48$ are included,
as indicated. The horizontal straight line shows the
Holtzer plateau $(= \pi)$ if the coarse-grained contour length
could be identified with the chemical contour length.}
\end{center}
\end{figure}

In previous work~\cite{11,17} we have already considered the 
decomposition of the total scattering function of bottle-brush 
polymers into the scattering from the backbone and from the side chains. 
This analysis which has the advantage that it provides a direct link to 
corresponding experiments~\cite{32,33} will not be addressed here, 
but we rather focus on the scattering function of the backbone only. 
Fig.~\ref{fig14} shows Kratky plots for relatively short backbone 
chain lengths ($N_b=131$ and 259, respectively). One recognizes that 
for short side chain lengths ($N_s=6,12)$ $qLS_b(q)$ does not settle down to a 
well-defined ``Holtzer Plateau'', at least not within the available 
window of wavenumbers. Clearly, also the range over which $q LS_b(q)$ 
decays from the maximum to the horizontal part that appears for 
$N_s=24$ and $48$ is rather small, and does not warrant any analysis 
in terms of the power laws suggested in Fig.~\ref{fig5}b. This mismatch 
between the actual plateau values (for $N_s=24$ and $48$), which are
close to $4$, and 
the theoretical value $\pi$ can be attributed to reduction of $L_{cc}$ in 
comparison to $N_b \bar{\ell}_b$ since the actual orientations of the 
backbone vectors are not strictly aligned with the coarse-grained 
backbone (Fig.~\ref{fig4}), as is also evident from the fact that 
$\langle \vec{a}_i \cdot \vec{a}_{i + 1} \rangle / \langle \vec{a}_i^2 \rangle$
(Fig.~\ref{fig2}b) is already reduced to about $0.7$, but the further 
decrease of $\langle \vec{a}_i \cdot \vec{a}_{i+s} \rangle 
/\langle \vec{a}_i^2 \rangle$ is rather slow, due to the side chain 
induced stiffening of the backbone on mesoscopic scales.
The ratio at about $4/\pi$ is compatible with the reduction of $L_{cc}$
by about $30\%$ relative to $L$ noted previously, so gratifyingly
our analysis is internally consistent.

\begin{figure}
\begin{center}
% for a multi-line caption
%\onelinecaptionstrue
% for a one-line caption
%\includegraphics{}
\includegraphics[scale=0.32,angle=270]{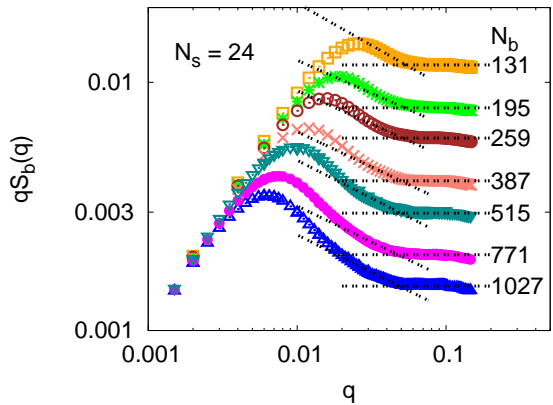}\hspace{0.4cm}
\caption{\label{fig15} Log-log plot of $qS_b(q)$ vs.~$q$ for bottle-brush
polymers with side chain length $N_s=24$ and various backbone chain lengths
$N_b$ from $N_b=131$ to $N_b=1027$, as indicated. The point of intersection
between two Broken straight lines illustrates the estimation of $q^*(N_b)$,
the wavenumber where the onset at the
Holtzer plateau occurs.}
\end{center}
\end{figure}

Fig.~\ref{fig15} shows plots of $S(q)$ vs.~$q$ for fixed side chain 
length $N_s$=24 but different backbone chain lengths. In this plot, 
an attempt is made to locate an onset wavenumber $q^*$ for the Holtzer 
plateau, in terms of a fit of two intersection straight lines. Of course, 
the data are smooth and the onset of the Holtzer plateau does not occur 
sharply but rather gradual; thus $q^*$ can be estimated only with 
considerable error (for large $N_b$ we estimate $q^* \approx 0.06 \pm 0.01$, 
while for $N_b=131$ the estimate rather is $q^*\approx0.075 \pm 0.020$~\cite{11}). 
Now the question is, how can one relate $q^*$ explicitly to the persistence 
length? Should one take $\ell_p= 2 \pi/ q^*$, or $\ell_p=1/q^*$? 
Lecommandoux et al.~\cite{28} who were the first to try such a method 
suggested the relation $\ell_p \approx3.5/ q^*$, but we see little 
theoretical support for this choice either.

It would be advantageous if one could rely on the des Cloizeaux relation, 
Eq.~(\ref{eq26}), which suggests to plot $qS(q)$ vs.~$1/q$ for 
$q \ell_p \gg 1$: one should find a straight line, the intercept at the 
ordinate should yield $\pi$, the slope of the straight line should yield $2/(3 \ell_p)$.

\begin{figure}
\begin{center}
% for a multi-line caption
%\onelinecaptionstrue
% for a one-line caption
%\includegraphics{}
(a)\includegraphics[scale=0.32,angle=270]{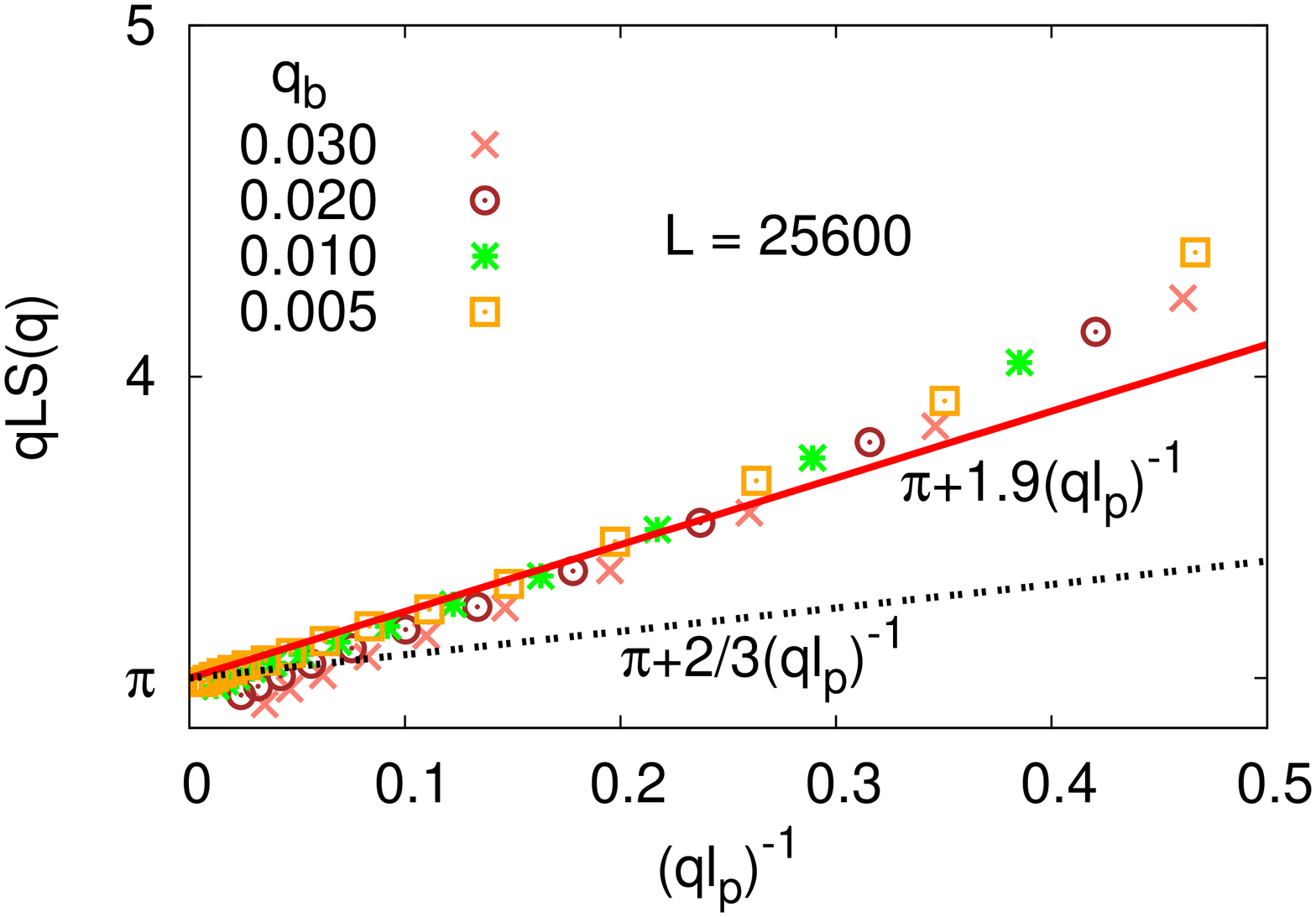}\hspace{0.4cm}
(b)\includegraphics[scale=0.32,angle=270]{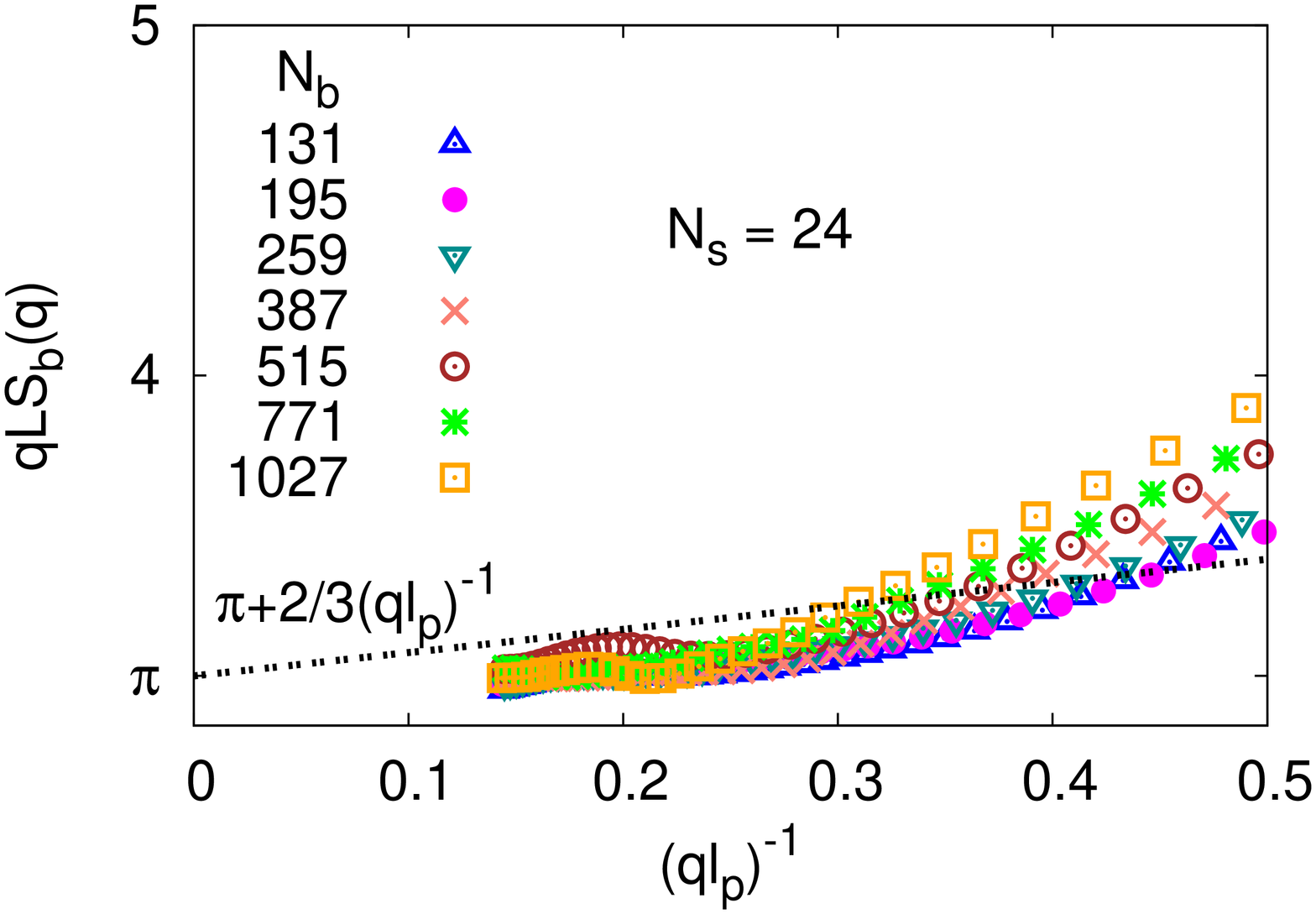}\\
\caption{\label{fig16} (a) Plot of $q L S (q)$ vs.~$(q \ell_p)^{-1}$
for the semiflexible SAW and $L=25600$, including 4 choices of $q_b$,
as indicated.
(b) Same as (a) for bottle-brush polymers with fixed side chain length
$N_s=24$ and varying backbone length as indicated.
Broken straight line is the des Cloizeaux~\cite{78} prediction,
Eq.~(\ref{eq26}), full straight line an empirical fit to the data.}
\end{center}
\end{figure}

However, when one tests this method for the semiflexible SAW, one finds that 
the data that can be fitted to a straight line are at $q \ell_p \approx 1$ 
rather than at $q \ell_p \gg 1$, and the slope of the straight line disagrees 
with the prediction (Fig.~\ref{fig16}a). Thus, it is not really a big 
surprise that this does not work well for our bottle-brush model 
either (Fig.~\ref{fig16}b).

\begin{figure}
\begin{center}
% for a multi-line caption
%\onelinecaptionstrue
% for a one-line caption
%\includegraphics{}
(a)\includegraphics[scale=0.32,angle=270]{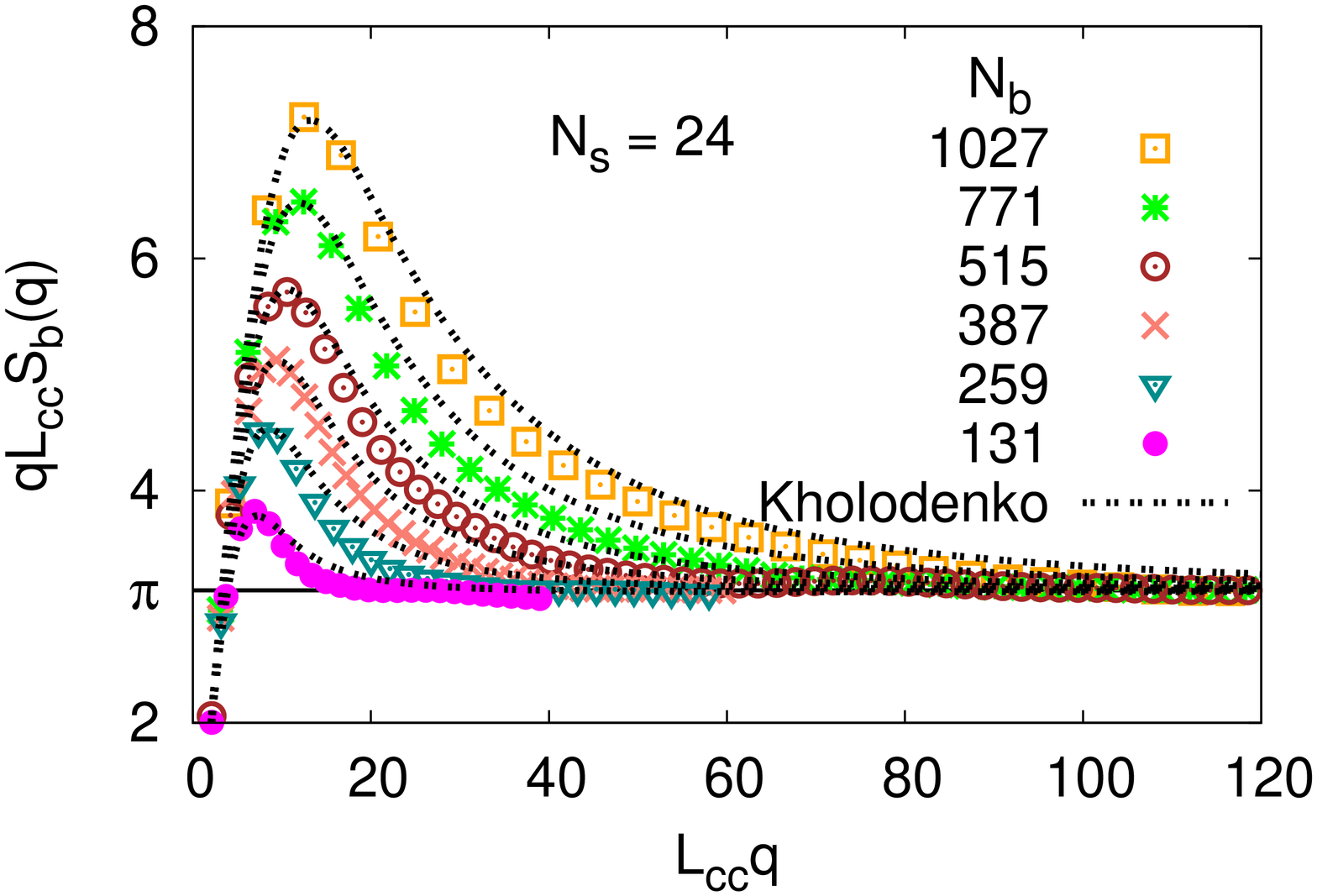}\hspace{0.4cm}
(b)\includegraphics[scale=0.32,angle=270]{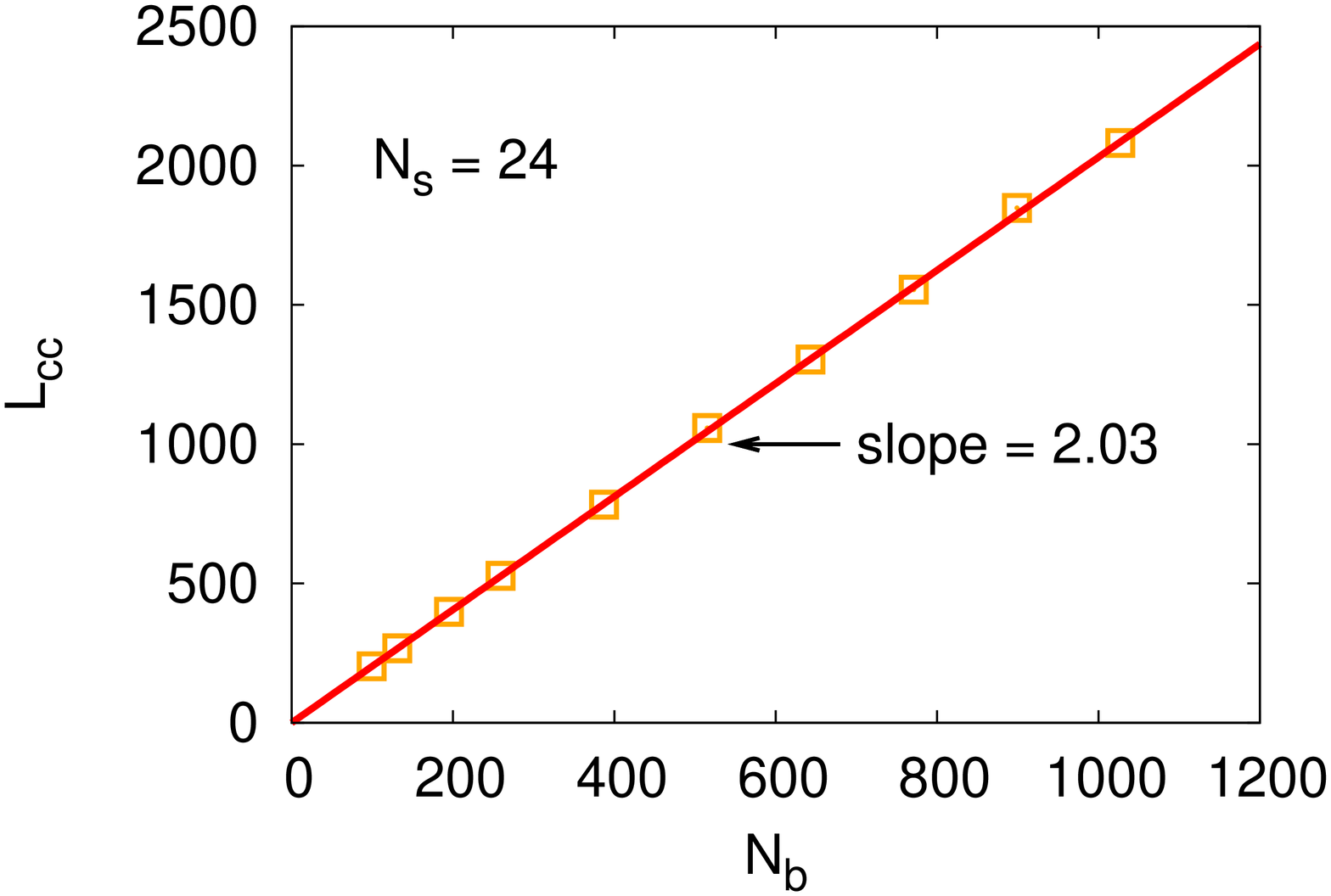}\\
(c)\includegraphics[scale=0.32,angle=270]{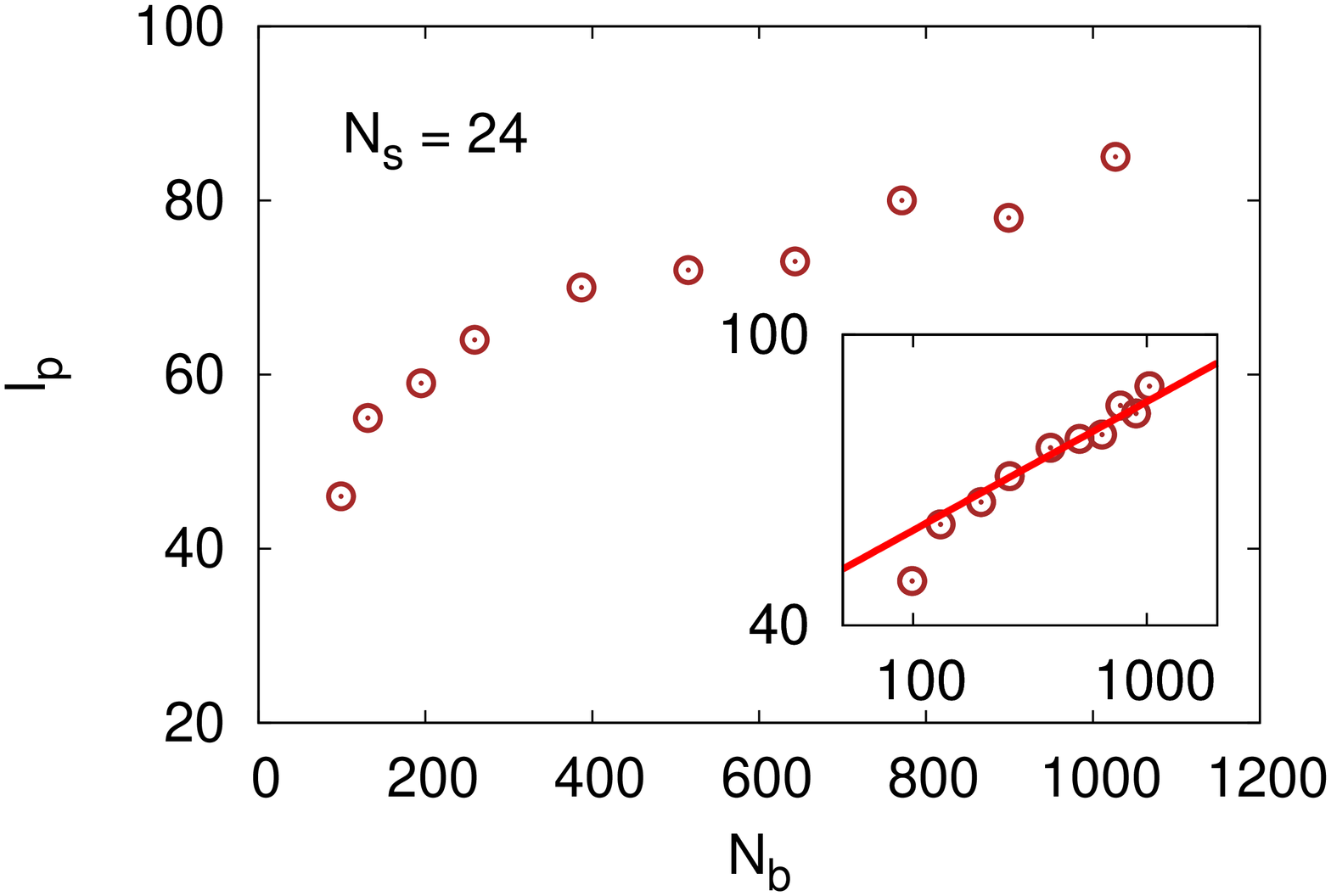}\hspace{0.4cm}
\caption{\label{fig17}
Plot of $q L_{cc} S_b (q)$ versus $L_{cc} q$ for bottle-brush polymer with
side chain length $N_s=24$ and several values of $N_b$, from $N_b=131$ to
$1027$, as indicated. For each choice of $N_b$ both $L_{cc}$ and $\ell_p$ were
individually adjusted. (b) Plot of $L_{cc}$ vs.~$N_b$, showing that
$L_{cc} \approx 2.03 N_b$ rather than being equal to the ``chemical''
contour length $L= \bar{\ell}_b N_b \approx 2.7 N_b$. (c) Plot of the
persistence length $\ell_p$ (from the fit in (a)) versus $N_b$.
In the log-log plot (inset), the theoretical power law
$\ell_p \propto N_b^{2\nu-1}$ is shown by the straight line.}
\end{center}
\end{figure}

An interesting alternative of data analysis is, however, a fit of 
the Kholodenko 
formulas \{Eqs.~(\ref{eq28})-(\ref{eq30})\} to the structure factor, 
using both $L_{cc}$ and $\ell_p$ as individual adjustable parameters for 
each value of $N_b$ (Fig.~\ref{fig17}). First of all, one sees that the 
Kholodenko structure factor provides a good fit in all cases, and the result 
for the coarse-grained contour length, $L_{cc}=2.03 N_b$, even is physically 
very reasonable: we have obtained that there is a 30\% reduction of 
the $L_{cc}$ in comparison to the ``chemical'' contour length 
$L=N_b \bar{\ell}_b=2.7 N_b$ in the previous subsection.

However, the problem of this fit is the unphysical behavior of the 
persistence length $\ell_p$: since we know that the Kholodenko~\cite{87} 
approach involves necessarily the Gaussian result 
$\langle R^2_g \rangle =\frac{1}{3} \ell_p L_{cc}$ but we know that for our 
model $\langle R^2_g \rangle \propto N^{2_\nu}_b$ and $L_{cc}\approx 2.03 N_b$, 
the only way to reconcile these results is a persistence length scaling as 
$\ell_p \propto N_b^{2 \nu-1}$, and this is what we see in Fig.~\ref{fig17}c.
Thus, despite the seemingly good fit (Fig.~\ref{fig17}a) and good 
results for $L_{cc}$ (Fig.~\ref{fig17}b), the results for the persistence 
lengths are completely unreliable!

In order to apply this approach, one must make sure that one works with data in the Gaussian regime, and this is not at all the case for bottle-brush polymers under good solvent conditions.

\section{Conclusions}

In this paper, we have focused on the behavior of single semiflexible polymers
under very good solvent conditions, considering how the chain stiffness affects
polymer properties such as the mean square gyration radius, the structure
factor, etc. Our analysis focused on the question how the variation of chain
stiffness affects these properties, and hence one can infer from these 
properties a characterization of the ``intrinsic stiffness'' of the
polymer chain in terms of the so-called ``persistence length''.

   We have contrasted two models, the self-avoiding walk on the simple
cubic lattice where a bending energy $\varepsilon_b$ causes pronounced 
stiffening of the polymer when $\varepsilon_b \gg k_BT$, and a lattice model 
for bottle-brush polymers, where backbone stiffening is 
caused by increasing the length
of side chains. These two models constitute two quite distinct limiting
cases: in the SAW model, increase of $\varepsilon_b/k_BT$ causes stiffening 
without any effect on the local thickness of the chain, which strictly 
remains the lattice spacing. For the bond fluctuation model of polymer 
brushes, however, we have found that backbone stiffening is caused by the
thickness of the (coarse-grained) worm-like chain, the persistence length
increases proportional to the cross-sectional diameter of the bottle-brush.

   Since snapshot pictures (Fig.~\ref{fig6}) suggest that the 
bottle-brush polymers (or their backbones, respectively) resemble 
worm-like chains (and the same conclusion is often drawn from AFM
pictures or electron micrographs of actual polymers), 
the use of the Kratky-Porod worm-like
chain model has become very popular. However, we demonstrate here that 
for bottle-brush polymers this model yields
very misleading results: since the
mean square gyration radii of bottle-brushes are found to scale with their
contour length $L$ as $\langle R_g^2 \rangle \propto L^{2\nu}$, the
Kratky-Porod (K-P) result $\langle R_g^2 \rangle = \ell_p L/3$ 
invariably causes a spurious contour length dependence of the persistence 
length when fit to the data, namely 
$\ell_p(L) \propto L^{2\nu-1} \rightarrow \infty$ as $L \rightarrow \infty$.
Although the fits of the K-P model look almost perfect (Fig,~\ref{fig17}a)
and numbers for $L$ resulting for the contour length from the fit are
rather reasonable, the result for ``the'' persistence length simply is 
meaningless!

    Already in our earlier papers we have shown that similar ambiguous
results for the persistence length are gotten when orientational
correlations along the chain backbone are analyzed, or the projection of
bond vectors on the end-to-end distance are studied (although the 
resulting numbers for $\ell_p(N_b)$ seem to be somewhat smaller
than those shown in Fig.~\ref{fig17}c). The large $q$-behavior of the
structure factor $S(q)$ yields a qualitatively more reasonable
behavior, but a unique choice for a well-defined persistence length as 
a measure for intrinsic chain stiffness does not emerge. All these 
difficulties in understanding the stiffness of bottle-brush polymers in good 
solvents are intimately linked to the fact that one can coarse-grain
into some effective self-avoiding walk model (Figs.~\ref{fig9}, \ref{fig10}),
and no regime exists where the polymers resemble Gaussian chains. Of course,
this fact is different if we would consider bottle-brush polymers
in Theta-solvents (as done by Theodorakis et al.~\cite{96}), since then
$\langle R_g^2 \rangle \propto L$ and the use of the K-P model is 
qualitatively reasonable. 
{Another interesting possibility to extract a persistence
length of bottle brushes would be an attempt to estimate an effective
bending modulues. One would have to estimate the coarse-grained free
energy of bent versus non-bent configurations of suitable pieces of
bottle-brush polymers, which in principle can be deduced from sampling
suitable angular distribution functions for such sub-chains. However,
the implementation of such an approach is not straightforward and has
not been attempted.}

    We have found that the situation in some respects is simpler if
one considers polymers where the stiffness can be enhanced while keeping 
their thickness constant, as modeled by a semiflexible extension of 
the standard SAW model. Then an intermediate Gaussian-like behavior of the
mean square radii and the structure factor emerges, and this can be 
understood theoretically (Fig.~\ref{fig5}), at least in qualitative terms. 
While still the asymptotic decay of bond vector autocorrelation 
functions is unsuitable to infer anything about the intrinsic stiffness
(due to the fact that the asymptotic decay is not exponential but rather
described by a power law), in favorable cases the initial decay of these 
autocorrelation
functions provided useful estimates of the persistence length, which then
can be used as input in the K-P model. While still some problems occur to 
understand for very long chain the crossover between the K-P model and the
ultimate SAW behavior, quantitatively, in qualitative terms the situation 
is understood. We emphasize, however, that all these comments only address 
the three-dimensional case: in $d=2$ dimensions, the K-P model
does not work at all, and one has a direct crossover from rod-like polymers
to SAW's.

   It is hoped that our analysis will help experimentalists with a 
proper interpretation of their data on semiflexible polymers.

\begin{acknowledgments}
We are grateful to the Deutsche Forschungsgemeinschaft (DFG) for support under
grant No SFB 625/A3, and to the John von Neumann Institute for Computing
(NIC J\"ulich) for a generous grant of computer time. We are particularly
indebted to S. Stepanow for his help with the explicit calculation of his
exact formula for the structure factor of the Kratky-Porod model. We are also
indebted to Hyuk Yu for pointing out Ref.~\cite{67} to us.
H.-P. Hsu thanks K. Ch. Daoulas for stimulating discussions.
\end{acknowledgments}

\end{document}